\documentclass[sigconf, nonacm]{acmart} 
\usepackage{times}
\usepackage{graphicx}
\usepackage{url,hyperref}
\usepackage{amsmath,amsfonts}
\usepackage{nicefrac}
\usepackage{algorithm}
\usepackage[noend]{algpseudocode}
\usepackage{xcolor}
\usepackage[utf8]{inputenc}
\usepackage[ngerman,english]{babel}
\usepackage{csquotes}
\usepackage{layout}
\usepackage{subfig}

\DeclareMathOperator{\enc}{Enc}
\DeclareMathOperator{\dec}{Dec}
\DeclareMathOperator{\kl}{KL}

\DeclareMathOperator{\softplus}{Softplus}

\begin{document}

\title{Learning Multiple-Scattering Solutions for Sphere-Tracing of Volumetric Subsurface Effects -- (\textsc{Preprint})}

\author{L.\,Leonard}
\affiliation{%
	\institution{University of Havana}
	\country{Cuba}}
\email{lleonart@matcom.uh.cu}

\author{K.\,H\"ohlein}
\affiliation{%
	\institution{Technical University of Munich}
	\country{Germany}}
\email{kevin.hoehlein@tum.de}

\author{R.\,Westermann}
\affiliation{%
	\institution{Technical University of Munich}
	\country{Germany}}
\email{westermann@tum.de}

\begin{teaserfigure}
	\includegraphics[width=\linewidth]{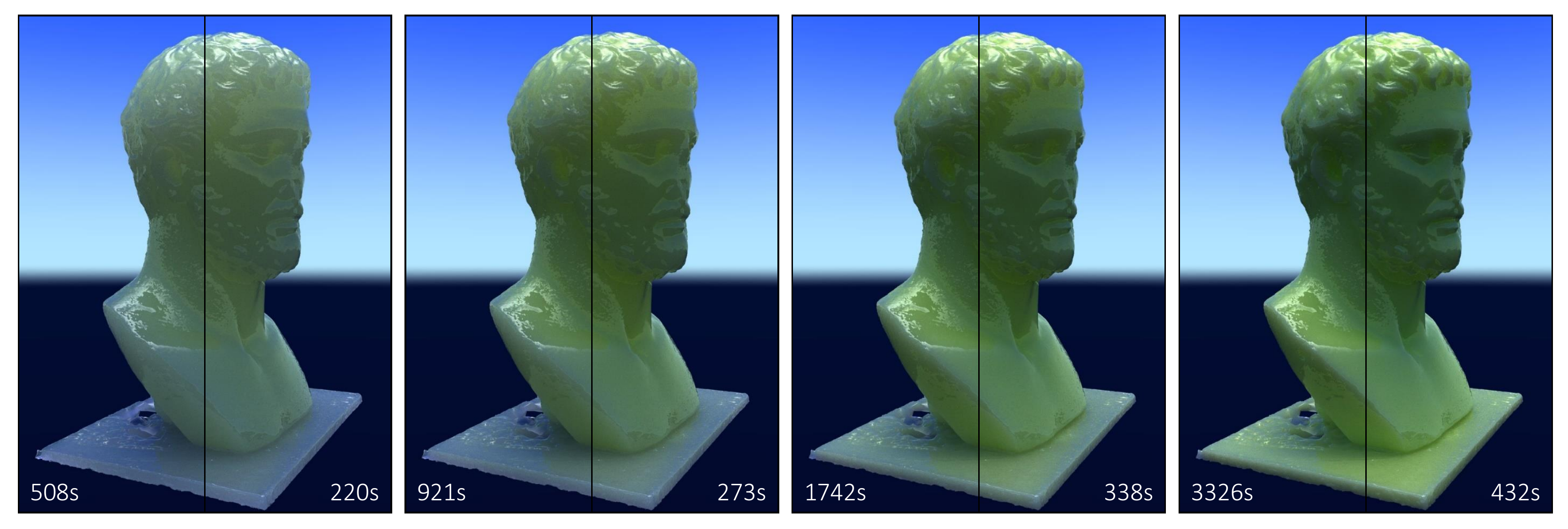}
	\centering
	\caption{Subsurface scattering simulation  for homogeneous media with varying density and high anisotropy ($g = 0.8$) using standard path-tracing (left) vs. our approach -- a sphere-tracing algorithm, based on data-driven learning of  multiple-scattering statistics (right). 5000 light paths per pixel are simulated. Density doubles from left to right, significantly increasing the number of scattering events to be considered during path-tracing. Our approach summarizes many scattering events into a single sphere-tracing step and thus minimizes performance loss due to exceedingly long scattering chains.}
	\label{fig:teaser}
\end{teaserfigure}

\begin{abstract}
	Accurate subsurface scattering solutions require the integration of optical material properties along many complicated light paths. We present a method that learns a simple geometric approximation of random paths in a homogeneous volume of translucent material. The generated representation allows determining the absorption along the path as well as a direct lighting contribution, which is representative of all scattering events along the path. A sequence of conditional variational auto-encoders (CVAEs) is trained to model the statistical distribution of the photon paths inside a spherical region in presence of multiple scattering events. A first CVAE learns to sample the number of scattering events, occurring on a ray path inside the sphere, which effectively determines the probability of the ray being absorbed. Conditioned on this, a second model predicts the exit position and direction of the light particle. Finally, a third model generates a representative sample of photon position and direction along the path, which is used to approximate the contribution of direct illumination due to in-scattering. To accelerate the tracing of the light path through the volumetric medium toward the solid boundary, we employ a sphere-tracing strategy that considers the light absorption and is able to perform statistically accurate next-event estimation. We demonstrate efficient learning using shallow networks of only three layers and no more than 16 nodes. In combination with a GPU shader that evaluates the CVAEs' predictions, performance gains can be demonstrated for a variety of different scenarios. A quality evaluation analyzes the approximation error that is introduced by the data-driven scattering simulation and sheds light on the major sources of error in the accelerated path tracing process.
\end{abstract}

\maketitle

\section{Introduction}


The realistic rendering of translucent solid materials such as wax or skin requires simulating the scattering of light in the interior of the body. Subsurface scattering refers to the mechanism where light that penetrates the boundary of a translucent object is scattered internally until it exits the body at a different location on the boundary. Due to the internal scattering of light and the resulting complicated light paths that have to be considered in the simulation, an accurate subsurface scattering solution becomes computationally very expensive. 

To reduce the computational requirements, approximation methods exploit special shape and material configurations, such as planar surface geometry, isotropic scattering, or separability of the bidirectional scattering-surface reflectance distribution function (BSSRDF) 
into location and direction dependent terms.
While it has been shown that efficient analytic transport solutions can be developed for some special cases, rendering quality decreases when the resulting algorithms are applied to scene configurations beyond the scope of the restrictive assumptions. To improve the generalizability of rendering acceleration solutions, there has been growing interest in data-driven approaches that can infer approximate multiple-scattering solutions at render time.

One of the earliest methods precomputes long-distance light transport through a homogeneous medium using transition probabilities at varying spatial scales~\cite{moon2007rendering}. This so-called shell tracing approach enables to move rays through the volume in large steps without considering individual scattering events. It builds upon the concept of sphere-tracing for surface rendering~\cite{hart1996sphere}, to adapt  the step size to how deep the ray is in the solid body. 
Recently, for rendering volumetric media without an explicit boundary representation, deep-scattering~\cite{kallweit2017deep} proposed to fed geometric information into a network by progressively evaluating a hierarchical volumetric geometry descriptor. Inference is based on an expressive multi-layer neural network. To address specifically the issue of boundary effects, a method for learning a shape-aware BSSRDF model from ground truth volumetric transport simulations using path tracing has been proposed by Vicini et al~\cite{vicini2019learned}. The method avoids many limiting assumptions of prior analytic models regarding the planarity of surfaces and isotropy of volumetric transport, and it generalizes the built-in notion of spatio-directional separability. It assumes a local polynomial surface expansion around a shading point, and uses this expansion to re-project predicted off-surface samples back onto the surface.

Our work builds upon prior works in data-driven subsurface scattering simulation, by leveraging the core idea of "bypassing a potentially lengthy internal scattering process" ~\cite{vicini2019learned} through a network that has learned to sample outgoing light locations on an object's surface from an incident light location. In contrast to existing work in subsurface scattering simulation, our method avoids an explicit encoding of the local surface geometry in the stochastic modelling of the scattering process, and it also infers the outgoing light direction from a reference distribution generated via volumetric path tracing, instead of using importance sampling. To achieve this, we combine shell tracing with network-based learning of long-distance light transport between a point and a spherical surface centered at this point. A sequence of neural-network-based conditional variational auto-encoders (CVAEs) is trained to model the statistical distribution of light paths and absorption along them inside a spherical region in presence of multiple scattering events. A first CVAE learns to sample the number of scattering events, i.e., the absorption, occurring along a path inside the sphere. Conditioned on this, a second model predicts exit position and continuation direction of this path. By using shell tracing, light paths can be traced efficiently through the volumetric medium toward the solid boundary without an explicit encoding of the boundary geometry.

Furthermore, instead of modelling purely the one-to-one light transport between two surface points, our method supports extensions of volumetric path tracing, including direct in-scattering to achieve improved convergence. For this, a third network model generates a representative sample of photon position and direction along the path, which is used to approximate the contribution of direct illumination due to in-scattering. In combination with existing approaches for simulating the direct interior illumination, the path-tracing solution converges quickly even when none of the paths arrives directly at a light source.

We jointly train all three networks on a spherical geometry with unitary radius and varying interior density (different radii can be considered by adjusting the density appropriately). In the training and validation phase, ground truth transport solutions using Monte Carlo path tracing are used.  
An interesting result is that all proposed networks are light-weight, comprised of only up to three layers with no more than 16 nodes each. Despite the resulting compact internal latent-space representations, very realistic reconstructions of the modeled distributions are obtained. Due to this compactness, all inferencing steps can be implemented efficiently using short shader programs and matrix multiplication instructions on the GPU. We analyze the performance and accuracy of our approach on different geometry and material properties, and we also investigate the inferencing skills of each of our network models individually. Our method achieves quality en par with that of a reference Monte Carlo path tracing solution, at significantly improved performance. 




\section{Related Work}

Compared to traditional ray tracing algorithms that can assume a free traverse of the light on the vacuum between surfaces, ray tracing participating media represents a superior level in computational complexity. For some medium such as milk and clouds, the high value of the scattering albedo (i.e.\ the ratio of energy remaining after scattering) entails a large number of light bounce simulations before the ray is absorbed, leaves the volume or hits a surface. Due to the plethora of possible material interactions, the rendering equation for realistic scene settings does not possess a closed-form solution in presence of participating media. Thus, for synthesizing realistic images, approximate solutions are required. 

Traditional approximation schemes include the exclusion of higher-order scattering effects, leading to the single-scattering approximation~\cite{sun2005practical} -- which under certain circumstances can admit a closed-form solution~\cite{pegoraro2009analytical} -- or approaches based on luminance diffusion~\cite{kajiya1984ray, farrell1992diffusion}. Especially approximations of the latter type, however, typically impose severe assumptions on homogeneity of the considered material and surface geometry, which effectively disables application of these approaches in more complex settings. Even increasing model capacity, for example by considering dipole diffusion~\cite{jensen2001practical}, does not resolve the shape restrictions. A more thorough review of traditional rendering techniques for participating media can be found in \cite{cerezo2005survey}

\emph{Monte Carlo methods} More flexible approaches arise from the application of Monte Carlo path tracing algorithms~\cite{Lafortune1996} for numerically solving the rendering equation for participating media \cite{kajiya1984ray,rushmeier1989realistic,hanrahan-krueger-1993}. Though being computationally expensive, Monte Carlo path tracing in a converged setting yields physically accurate images which can serve as a "ground-truth" for validating alternative approaches. The recent survey by Novak et al. \cite{novak2018monte} reviews the latest advances in Monte Carlo path tracing methods for solving the light transport in participating media. 

In path tracing and as an alternative to ray-marching with constant step size, delta tracking is often used to determine free path lengths according to the optical depth in the participating medium. As an importance sampling of the corresponding probability density function, Woodcock tracking \cite{woodcock1965techniques} adjusts the sampling distances so that dense regions in a volume can be sampled appropriately, and it has been adapted to achieve unbiased sampling of sparse inhomogeneous media \cite{yue2010unbiased}. Free path sampling with probabilities not necessarily proportional to the volume transmittance has been realized by means of weighted delta tracking approaches, e.g.~\cite{novak2014residualtracking,kutz2017tracking,rehak2019weighted}. All these techniques reduce noise in the estimated light paths for participating media, and they enable an automatic selection of the step size according to the underlying material distribution. 

A different approach to move rays through the volume in adaptive steps even without considering individual scattering events is shell tracing ~\cite{moon2007rendering}. It builds upon a statistical model that is based on a discretized array of transition probabilities, limiting the expressiveness of the simulated light transport paths. In the spirit of sphere-tracing for surface rendering~\cite{hart1996sphere}, it performs larger step in the interior of a body and successively decreases the step size towards the object boundary. 

\emph{Diffusion methods} Despite the large generality of Monte Carlo Rendering, a variety of methods have been developed, which are based on diffusion approximations. Especially in the field of sub-surface scattering, computation times can explode when highly-dense materials require translucent renderings. Though the light does not enter the bodies deeply, Monte-Carlo approaches may run into lengthy sequences of frequent scattering or reflection events right underneath the surface. Diffusion models, which approximate the light transport by virtue of bidirectional scattering-surface reflectance distribution functions (BSSRDFs) may in these cases lead to speed-ups. While early approaches, like~\cite{jensen2001practical}, can hardly keep up with the quality of ray-traced images, more recent methods like~\cite{d2011quantized,yan2012accurate,frisvad2014directional} can generate visually appealing results at reduced rendering time, however, at the cost of reduced physical realism. Notably, also hybrid methods exist, which weight out the advantages and disadvantages of Monte Carlo-based and diffusion-based rendering against each other~\cite{habel2013photon}. 





\emph{Machine Learning for Rendering}
As Monte Carlo-based rendering can intrinsically be described as a statistical problem, the use of machine learning methods for rendering has recently attracted great interest. In the context of path tracing the goal is typically to use machine learning models to model sampling distributions or create guidance for efficient sample selection. Popular modeling approaches include Gaussian mixture models~\cite{vorba2014line,herholz2016product}, and neural network methods. 

\emph{Neural network-based methods} In the context of path tracing, deep learning methods can be classified into image-based adaptive sampling and denoising algorithms on the one hand, and multiple-scattering approximations on the other hand. 

The recent approach by Kuznetsov et al.~\cite{kuznetsov2018deep} facilitates neural networks for learning adaptivity in Monte-Carlo path-tracing and denoising of the final image. A first network learns to adapt the number of additional paths from an initial image at the target resolution, which is generated via one path per pixel. A second denoising network learns to model the relationship between an image with increased variance in the color samples to the ground truth rendering ~\cite{cnn-denoise,mara17towards}. Recently, Weiss et al.~\cite{weiss2020learning} used neural networks to learn the positions of sample locations in image space from a low resolution image, using a differentiable sampling stage as well as a differentiable image reconstruction stage that can work on a sparse set of samples. Related to denoising approaches for path traced images is the screen space shading approach by Nalbach et al. ~\cite{nalbach2017deep}, where a network is trained to infer missing shading information from image information in a view-dependent G-buffer. 


For surface graphics, Zheng and Zwicker have used neural networks to model the relationships between scene parameters and light paths ~\cite{zheng2019learnsample}. Recently, deep-scattering~\cite{kallweit2017deep} employed a radiance-predicting neural network for simulating scattering events in clouds. Geometry information is fed into the network by progressively evaluating a hierarchical volumetric geometry descriptor, and inference is based on an expressive multi-layer neural network, optimized with recent design features. Nevertheless, the method was developed for cloud rendering and does, as such, not explicitly account for boundary effects, which may play a major role in sub-surface scattering. Adrressing specifically this issue, a method for learning a shape-aware BSSRDF model from ground truth volumetric transport simulations using path tracing has been proposed~\cite{vicini2019learned}. Exploiting the expressiveness of neural network models in a similar way, this method avoids many limiting assumptions of prior analytic models regarding the planarity of surfaces and isotropy of volumetric transport, and it generalizes the built-in notion of spatio-directional separability. It assumes a locally expandable surface geometry, which is explicitly encoded using a trivariate polynomial around a shading point. In this way, an approximate signed-distance function to the surface geometry can be used to model the outgoing on-surface light distribution. 

In \cite{vicini2019learned}, a learned approximation of all the subsurface scattering allows them to overcome the limitation of the planar approximation, fitting the geometry with a cubic polynomial. Nevertheless, their assumption of the final outgoing direction with a uniform distribution and a single polynomial approximation to the whole geometry biases the final result.

Another idea, considering the light scattering in a medium as a random process (similar to the walk-on-sphere statistic technique \cite{muller1956some}\cite{kyprianou2018unbiased}) is to have a short-cut of all possible scattering inside a sphere with homogeneous density. This model can be used to simulate all those ``events'' as a single step in a Montecarlo technique. This strategy represents an acceleration rather than an approximation, and removes any bias regarding the geometry, limitation of other techniques. In \cite{moon2007rendering} they propose a method named ``shell-tracing'' dealing with the rendering of discrete random media using precomputed scattering distributions in form of matrix factorization. With this approach, the number of variables considered is reduced to three due to the increment of the matrix complexity if a higher dimension is considered.

\section{Background and methods}
\label{sec:svs}

\begin{figure}
	\centering
	\includegraphics[width=\columnwidth]{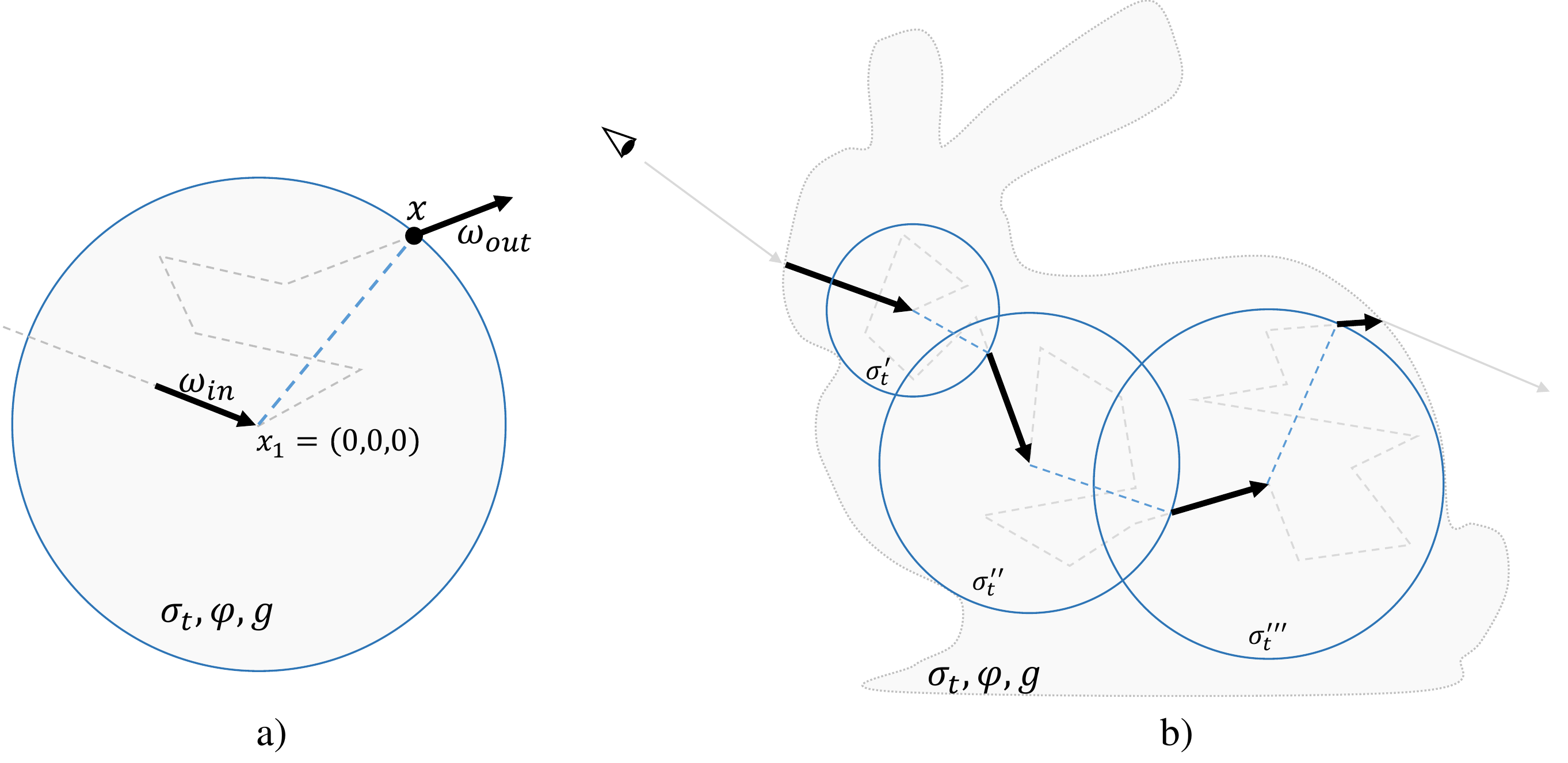}
	\caption{Method overview. a) Given optical properties of the volumetric medium ($\sigma_t,\varphi,g$), the path (gray dotted lines) starting at $x_1=(0,0,0)$ from the incoming direction $w_1=w_{\textnormal{in}}$ is bypassed by letting a network infer directly (blue) the random outgoing position $x$ and direction $w_{\textnormal{out}}$ in an unitary radius sphere. b) After free-travel events (black arrows), from a signed distance field the maximum sphere not intersecting the geometry is computed, and the radius is used to control the adaptive density and the length of the next step along a light path. This process is repeatedly applied until the path leaves the volume.}
	\label{fig:sphereScattering}
\end{figure}

Our key idea is to train generative statistical models to bypass multiple scattering events occurring inside a spherical volume of material with constant density and potentially anisotropic scattering characteristics, and to use the learned representations in the rendering process to generate paths through the volume using large steps (Fig.~\ref{fig:sphereScattering}a). Whenever the ray arrives at a sampling point, it tests for the largest step it can make without leaving the volume, and then evaluates a sequence of statistical models to infer the end position and direction when performing this step. 
The ray proceeds using delta tracking until a next collision event is determined, and recurrently performs learning-based path tracing (Fig.~\ref{fig:sphereScattering}b). To determine the length of the step, a 3D signed distance function is computed for the current geometry in a pre-process. For this, the object's bounding volume is discretized using a 3D voxel grid, and for each voxel a conservative shortest Chebyshev distance to the object's surface is computed \cite{deakin2019accelerated}. 
This representation enables to quickly find the radius of the sphere that can be safely traversed without intersecting the geometry. As the ray approaches the edges, it is forced to take smaller and smaller steps and eventually the tracking will pass across the boundary surface. To efficiently trace against a polygonal boundary surface, we employ a GPU ray-tracer using NVIDIA's RTX ray tracing interface~\cite{NVIDIA2018} through the DirectX Raytracing API (DXR).\

\subsection{Scattering model and data generation}

Learning is conducted in an offline process. We assume homogeneous extinction $\sigma_\text{t}$ and scattering albedo $\varphi$ (with $\sigma_\text{t}=\sigma_\text{a}+\sigma_\text{s}$ considering the rate of photons getting absorbed $\nicefrac{\sigma_\text{a}}{\sigma_\text{t}}$ or scattered into other directions $\varphi=\nicefrac{\sigma_\text{s}}{\sigma_\text{t}}$ due to light-particle interaction). 
Anisotropy in scattering is modelled by using the Henyey-Greenstein~\cite{henyey1941diffuse} phase function,
\begin{equation}
\rho_g(\omega',\omega) = \frac{1}{4\pi}\frac{1-g^2}{(1 + g^2 - 2g \cos(\omega,\omega'))^{3/2}}
\end{equation}
with $\cos( \cdot, \cdot )$ indicating the cosine between the respective directions and anisotropy parameter $g$, which equals the expected cosine between ray direction before and after scattering. For instance, $g > 8$ results in a strong forward lobe of scattering into the incoming light direction. In this setting, the radiative transport \cite{chandrasekhar1950radiative} can be described in terms of the volume rendering equation ~\cite{kajiya1984ray},
\begin{align}
\label{eq:vre}
L(x,\omega) &= \int_{t=0}^{d_z} T(x, x_t)\left[ \sigma_\text{a}\, L_\text{e}(x_t, \omega) + \sigma_\text{s}\,L_\text{s}(x_t,\omega)\right] \text{d}t\nonumber\\
&\quad+ T(x, z)\, L_\text{d}(z, \omega)\text{,} \nonumber
\end{align}
with $ x_t = x -\omega \,t$ and $z=x_{d_z}$. $L(x, \omega)$ describes the incoming radiance at position $x$ from direction direction $\omega$, $L_\text{e}(x, \omega)$, $L_\text{s}(x, \omega)$ and  $L_\text{d}(x, \omega)$ denote emitted, scattered and direct radiance, and $T(x, y)$ is the transmittance along the ray (Beer-Lambert law \cite{lambert1760photometria}), with $T(x, x_t) = \exp(-\sigma_\text{t}t)$ in the case of homogeneous medium. The impact of the phase function is considered in $L_\text{s}(x_t, \omega)$, which can be rewritten as
\begin{equation*}
L_\text{s}(x_t, \omega) = \int_\Omega \rho_g(\omega, \omega')\,L_\text{i}(x_t, \omega')\,\text{d}\omega'\text{,}
\end{equation*}
where $L_\text{i}(x_t, \omega')$ denotes the directional incoming radiance at $x_t$, and the integral runs over the full solid angle $\Omega$. For a detailed discussion of the derivation of the rendering equation, we refer the reader to \cite{pharr2016physically}.

For training our model, we assume that rays start in the center of a sphere of unitary radius, and terminate when they cross the boundary of the sphere for the first time, after experiencing a sequence of scattering events in between. By assumption, the incident direction of the rays in the sphere center is set to $\omega_\text{in}^{\text{CVAE}} = (0, 0, 1)$ for all rays, and the first scattering event occurs directly in the center of the sphere. For each ray, 
Monte Carlo path tracing is used as a ground truth renderer to simulate radiative transport through scattering media ~\cite{Lafortune1996,novak2018monte}.
The 
distance between subsequent scattering events is estimated as $l = -\log(1 - \xi)/\sigma_\text{t}$, wherein $\xi$ is a uniformly distributed random number between $0$ and $1$. I.e., the mean free path is equal to the reciprocal of the extinction.  
At every scattering event, it is decided whether the path continues or terminates due to absorption using importance-based Russian roulette~\cite{eckhardt1987monte}. In case of continuation, the scattering direction is computed by importance sampling the Henyey-Greenstein phase function.

Within our model, we ignore the details of the full path trace. Instead, we assume that -- in the simplest case of a standard path tracer -- the only states which affect the outcome of the rendering are the number $N$ of scattering events affecting a ray inside the sphere, determining the likelihood of absorption $A$ (ray terminates due to absorption, $A = 1$, or not, $A = 0$), as well as the exit position $x$ on the shell of the spherical volume and the outgoing direction $w$ of the ray. To further support the use of the proposed method in the more general context of path tracing, we include the opportunity to sample additional ray statistics, which we summarize as $\Sigma$.

Following up these considerations, the likelihood of sampling a particular ray path with summary statistics $(N, A, x, \omega, \Sigma)$ is determined by a conditional probability density function (PDF) of the form $p(N, A, x, \omega, \Sigma | \sigma_t, g, \varphi)$, wherein quantities to the right of the vertical bar affect the PDF only as conditions. 

\subsection{Model architectures and inference\label{sec:cvaedescribtion}}

Empowered by fast progress in machine learning, in particular deep learning, a variety of approaches have been proposed to efficiently learn complex statistical distributions. These include variational auto-encoders \cite{kingma2013auto}, generative adversarial networks~\cite{goodfellow2014generative} or invertible neural networks~\cite{ardizzone2018analyzing}, which all can be extended to learn conditional distributions, see e.g.~\cite{NIPS2015_5775, mirza2014conditional, ardizzone2019guided}.
Within this study, we propose to train conditional variational auto-encoders (CVAEs) to learn a representation of $p(N, A, x, \omega, \Sigma | \sigma_t, g, \varphi)$, which offer a preferable trade-off between computational model complexity, model flexibility and training stability.

\begin{figure}
	\centering
	\subfloat[Working principle of CVAE.\label{fig:cvae}]{    \includegraphics[width=.48\columnwidth]{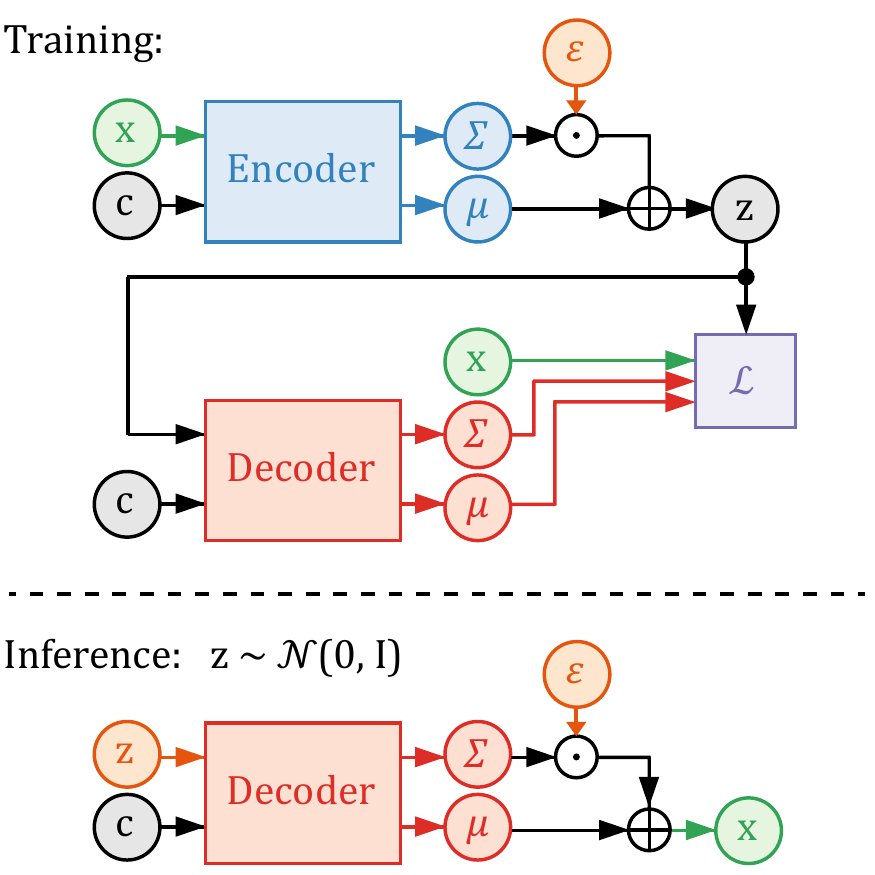}
	}
	\hfill
	\subfloat[Model realization in this work.\label{fig:ende}]{
		\includegraphics[width=.48\columnwidth]{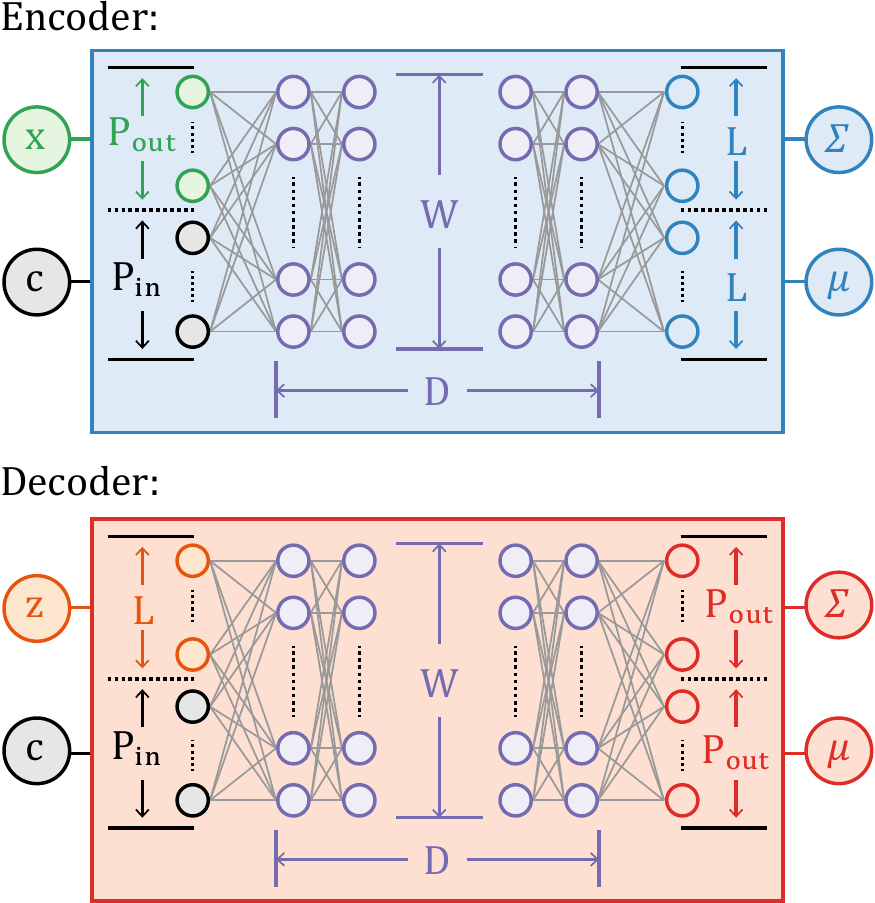}
	}
	\caption{(a) Working principle of the CVAE. 
		Parameters $\epsilon$ are random variables to evaluate the Gaussian posterior distributions, parametrized by mean $\mu$ and covariance $\Sigma$ of the respective model. (b) Configuration of encoder (blue) and decoder (red) networks for CVAEs. Both models are determined through the number of input parameters, $P_\text{in}$ (dimension of the conditioning variables, $c$), the number of output parameters, $P_\text{out}$ (dimension of the variables to be predicted, $x$), the latent space dimension $L$, the depth of the networks $D$, and the number of nodes per layer $W$. Encoder and decoder share the same settings of $D$, $W$ and $L$, and output estimates for mean $\mu_\text{Enc/Dec}$ and diagonal covariance matrix $\Sigma_\text{Enc/Dec}$ of the respectively modeled distributions.}
\end{figure}

In general, a CVAE defines a directed graphical model, which learns to emulate complex conditional data distributions, $p_\mathcal{X}(x \mid c)$, with data $x\in \mathcal{X}$ and condition $c\in\mathcal{C}$, by establishing a mapping between the original data space $\mathcal{X}$ and a latent feature space $\mathcal{Z}$. In this feature space, the data samples follow a prior distribution, $p_\mathcal{Z}(z \mid c)$, $z\in\mathcal{Z}$, from which samples can be drawn more easily, such as a multi-variate Gaussian distribution $\mathcal{N}(0, \mathbb{I}_\mathcal{Z})$ with zero-mean and diagonal unit variance. The distributions are related via
\begin{align*}
p_\mathcal{X}(x\mid c) &= \int_{\mathcal{Z}} p_\text{Dec}(x \mid z, c) p_\mathcal{Z}(z \mid c)\, \text{d}z\text{,} \\
p_\mathcal{Z}(z\mid c) &= \int_{\mathcal{X}} p_\text{Enc}(z \mid x, c) p_\mathcal{X}(x \mid c) \,\text{d}x\text{,}
\end{align*}
which can be combined to yield a self-supervised learning model, similar to the standard auto-encoder. Therein $p_\text{Enc}$ and $p_\text{Dec}$ denote probabilistic encoding and decoding mappings between data and latent space. Exploiting these relations, the CVAE defines a pair of neural network models, the encoder $\enc(x, c)$ and the decoder $\dec(z, c)$, which learn to parameterize the encoding and decoding PDFs $p_\text{Enc}$ and $p_\text{Dec}$. While the data-space posterior $p_\text{Dec}$ is typically considered to reflect the true mapping between $\mathcal{Z}$ and $\mathcal{X}$, the latent-space posterior $p_\text{Enc}$ can only be treated approximately, since the marginalization required to derive the exact latent-space posterior is computationally unfeasible. Within our study, we assume a conditional Gaussian shape for both distributions, i.e.,
\begin{align*}
q_\text{Enc}(z \mid x, c) &=  \mathcal{N}(\mu_\text{Enc}(x, c), \Sigma_\text{Enc}(x, c))\text{,}\\
p_\text{Dec}(x \mid z, c) &= \mathcal{N}(\mu_\text{Dec}(z, c), \Sigma_\text{Dec}(z, c))\text{,}
\end{align*}
wherein $q_\text{Enc}(z \mid x, c)$ refers to the parametric approximation of $p_\text{Enc}(z \mid x, c)$. Furthermore, $\mu_\text{Enc}(x, c)$ and $\Sigma_\text{Enc}(x, c)$, as well as
$\mu_\text{Dec}(z, c)$ and $\Sigma_\text{Dec}(z, c)$ are outputs of the encoder model $\enc(x, c)$, and decoder model $\dec(z, c)$, respectively, with $\mu_\text{Enc}(x, c) \in \mathcal{Z}$, $\mu_\text{Dec}(z, c) \in \mathcal{X}$. $\Sigma_\text{Enc}$ and $\Sigma_\text{Dec}$ are diagonal covariance matrices, consistent with the dimensions of $\mathcal{Z}$ and $\mathcal{X}$, respectively.
At training time, encoder and decoder models are trained jointly to maximize the so-called evidence lower bound
\begin{equation*}
\mathcal{L} = -\kl(q_\text{Enc}(z\mid x, c) \mid\mid \mathcal{N}(0, \mathbb{I})) + \langle \log p_\text{Dec}(x \mid z, c) \rangle_{q_\text{Enc}}\text{,}
\end{equation*}
which constitutes a lower bound for the observed data log-likelihood~\cite{kingma2013auto}. The first term refers to Kullback-Leibler divergence \cite{kullback1951information} between the learned latent-space posterior and the latent-space prior, and penalizes a mismatch between both distributions. The second term denotes the expected log-likelihood of the learned data posterior, where the expectation is taken with respect to the latent-space posterior, and penalizes the reconstruction error of the model. At inference time, the encoder model is discarded and a latent-space sample $z$ is drawn from the latent-space prior distribution $\mathcal{N}(0, \mathbb{I}_\mathcal{Z})$. Conditioned on this sample, the decoder model is evaluated and a data-space sample is obtained as
\begin{equation*}
x = \sqrt{\Sigma_\text{Dec}(z, c)}\,\epsilon + \mu_\text{Dec}(z, c)\text{,}
\end{equation*}
where $\epsilon$ is another Gaussian random variable, drawn from $\mathcal{N}(0, \mathbb{I}_\mathcal{X})$. This so-called reparameterization trick~\cite{kingma2013auto} assures that $x\sim\mathcal{N}(\mu_\text{Dec}, \Sigma_\text{Dec})$. A sketch of the training and inference processes in the current setting is shown in Figure~\ref{fig:cvae}.

In our work, we realize CVAE models via pairs of neural networks, encoder and decoder, respectively. During rendering, only the decoder is evaluated and, thus, needs to be kept as simple as possible. All our models follow the scheme shown in Figure~\ref{fig:ende} and are determined through the number of input parameters $P_\text{in}$, the number of output parameters, $P_\text{out}$, the latent space dimension $L$, the number of layers, i.e., the depth of the networks $D$, and the number of nodes per layer $W$. Both encoder and decoder networks use Softplus activation functions~\cite{zheng2015improving},
\begin{equation*}
\softplus(x) = \log(1 + \exp(x))\text{,}
\end{equation*}
between hidden layers. The outputs of the final layers, mean $\mu_\text{Enc/Dec}$ and diagonal covariance matrix $\Sigma_\text{Enc/Dec}$ for the respective parameters, are not transformed by non-linear activation functions. Note however, that for standard deviations we found it useful to predict $\log(\Sigma_\text{Enc/Dec})$ instead of $\Sigma_\text{Enc/Dec}$ directly. Both encoder and decoder networks share the same settings for $D$ and $W$. For a given set of input and target parameters, we examined different settings for $D$, $W$ and $L$, and empirically selected those parameters resulting in the most successful models. Parameter selection was guided by considerations with respect to computational efficiency. Since matrix multiplications on the GPU are carried out in matrix sections of size $4\times 4$, we adapted $D$, $W$ and $L$ so that especially the decoder model is able to efficiently utilize the parallelization capabilities on the GPU. In particular, we considered multiples of four for $W$-values, to speed-up matrix multiplications between hidden layers, and selected $L$ so that the input size of the decoder, $P_\text{in} + L$, is a multiple of four. Concerning $D$, we considered small values, as deeper models require longer computation time due to the sequential evaluation of subsequent layers. The models were trained using an AdamW optimizer~\cite{loshchilov2017decoupled}, implemented in the PyTorch interface in Python. Once the training was completed, the models were re-implemented in a shader-viable form, using mainly matrix multiplications and the nonlinear activation functions.

\begin{figure}
	\centering
	\includegraphics[width=.48\columnwidth]{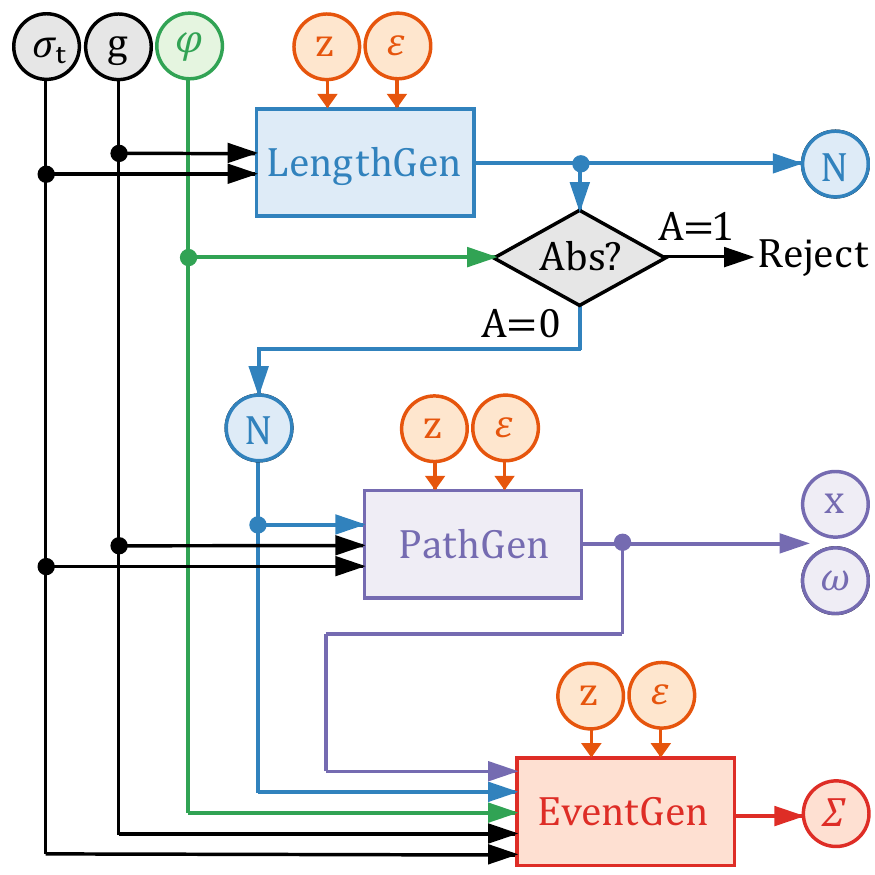}
	\caption{Overview of the overall model architecture, consisting of three trainable CVAE models, \textsc{LengthGen}, \textsc{PathGen}, \textsc{EventGen}, and an absorption request. Variables $z$ and $\epsilon$ refer to random numbers, which are used for sampling the generative models.}
	\label{fig:fullmodel}
\end{figure}

Recall now that the likelihood of sampling a particular ray path with summary statistics $(N, A, x, \omega, \Sigma)$ is determined by a conditional PDF of the form $p(N, A, x, \omega, \Sigma | \sigma_t, g, \varphi)$. To enable sequential model evaluation and minimize the need for modeling complex correlation structures between random variables, we decompose the PDF according to 
\begin{align}
p(N, A,  x, \omega, \Sigma | \sigma_\text{t}, g, \varphi) &= p_N(N \mid \sigma_\text{t}, g) \label{eq:probabilities}\\
&\quad\times p_A(A\mid N, \varphi) \nonumber \\
&\quad\times p_{x, \omega}(x, \omega \mid N, A, \sigma_\text{t}, g) \nonumber\\
&\quad\times p_{\Sigma}(\Sigma \mid N, A, x, \omega, \sigma_\text{t}, g, \varphi)\nonumber\text{,}
\end{align}
and decompose the full model into three separate CVAEs, which we will call \textsc{LengthGen}, \textsc{PathGen} and \textsc{EventGen} (Figure~\ref{fig:fullmodel}), one for each of the distributions $p_N$, $p_{x,w}$ and $p_\Sigma$. At inference time, the models are evaluated sequentially. The first model then infers the number $N$ of scattering events on the path. The PDF $p_N$, thereby depends only on $\sigma_\text{t}$ and $g$, because we suppose in this step that a full path is observed, which starts in the center of the sphere and terminates when crossing the sphere boundary without being absorbed in between. This approach simplifies the generation of training data, especially in scenarios with significant absorption. Conditioned on $N$ and $\varphi$, the probability of absorption $A$ along the bypassed path follows a Bernoulli distribution 
\begin{equation*}
p_A(A \mid N,\varphi) = q(N,\varphi)^A (1-q(N,\varphi))^{1-A}\text{,}
\end{equation*}
with $q(N,\varphi) = 1 - \varphi^N$. This admits simple sampling, so that training of a CVAE is not required. At this stage, a path can be rejected, if sampling yields $A = 1$. In this case, no further models have to be evaluated, increasing computational efficiency. 

If the path is accepted, the second CVAE infers the position on the sphere surface where the ray exits the volume as well as the direction into which it continues to be propagated (Figure~\ref{fig:sphereScattering}a). Note here that $p_{x,\omega}$ is independent of $\varphi$, which can be assumed since the absorption process is treated separately, before $x$ and $\omega$ are sampled, and $x$ and $\omega$ are conditioned on the outcome of the absorption request $A$. If applicable, the third model finally resolves the summary statistics $\Sigma$. Since in general no dependencies can be excluded, $p_\Sigma$ may depend on all previously sampled variables, as well as on all material parameters. An application of these statistics will be discussed in Section~\ref{sec:directillum}, where $\Sigma = (X, W)$ refers to a representative sample of all scattering events along the bypassed path inside the sphere. In this setting, $X$ denotes the position of the representative scattering event and $W$ the in-going direction of the photon path at $X$. 

\section{Learning multiple scattering in spherical volumes\label{sec:ls}}

\begin{figure}[h]
	\centering
	\includegraphics[width=0.7\columnwidth]{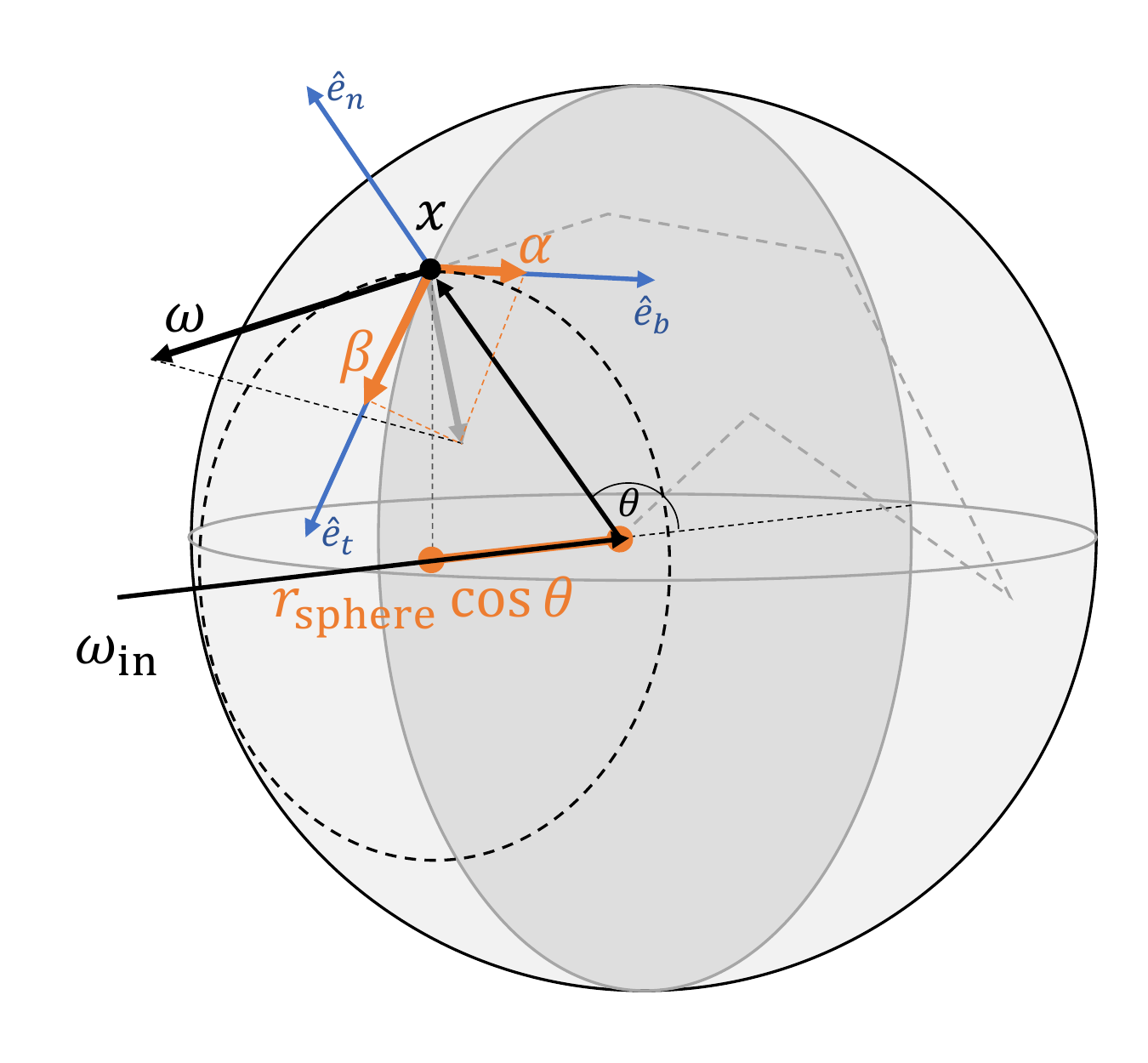}
	\caption{Path parameterization. Due to rotation symmetry of the phase function, it is useful to parameterize exit position $x$ and outgoing direction $\omega$ of a ray invariant with respect to rotations around $\omega_\text{in}$. All exit positions on a circle of constant deviation angle $\theta$ with respect to $\omega_\text{in}$ (dotted circle) are equally likely. Exit position $x$ of a path is therefore parameterized through $\cos(\theta)$. Coordinates of the outgoing direction $\omega$ are given in a local reference frame around $x$, with normal vector $\hat{e}_\text{n}$, binormal $\hat{e}_\text{b}$ and tangent $\hat{e}_\text{t}$, yielding coordinates $\alpha$ (projection on $\hat{e}_\text{b}$) and $\beta$ (projection on $\hat{e}_\text{t}$).
	}
	\label{fig:pathgen}
\end{figure}

While training of the CVAEs can be performed in normalized conditions, i.e.\ in a unit sphere with all rays coming from the same direction, the parameters have to be adapted to the local frame of reference of the current ray at render time. Also, the radius of the sphere has to be adapted to the geometry of the scene setting. To account for this, the output parameters of the models, i.e., $x$, $\omega$, $X$ and $W$, have to be parameterized in a way that is invariant to rotations of the spatial coordinates and changes of the sphere radius. To account for this, we propose a parameterization scheme as shown in Figure~\ref{fig:pathgen}.

\subsection{Path parameterization\label{sec:pathparam}}

After sampling $N$ and $A$, the outgoing position $x$ on a sphere shell of radius $r_\text{sphere}$ is parameterized 
in spherical coordinates through a tuple $(r_\text{sphere}, \theta, \psi)$, where $\cos(\theta) = \cos(\omega_\text{in},x)$. Due to invariance of the distribution $p_{x,\omega}$ with respect to rotations around $\omega_\text{in}$, $\psi$ can be sampled as a random number from a uniform distribution between $0$ and $2\pi$. This invariance is inherited from the rotation invariance of the Henyey-Greenstein phase function $\rho_g(\omega, \omega')$ and can be used to restrict the learning-space of the CVAE to only the $\theta$-coordinate. The rotation invariance furthermore simplifies the matching between real-space coordinates and CVAE reference system, since the orientation of the basis vectors orthogonal to $\omega_\text{in}$ can be chosen arbitrarily, as long as the CVAE-reference direction $\omega_\text{in}^{\text{CVAE}} = (0,0,1)$ is mapped appropriately to the true $\omega_\text{in}$ in real-space coordinates. Given a suitable rotation matrix $R^\text{CVAE}$, the direction $x$ can be sampled as
\begin{equation}
x = r_\text{sphere} R^\text{CVAE}R_\psi^{(0,0,1)} \tilde{x}\text{,}
\label{eq:transform}
\end{equation}
where $\tilde{x} = (\sin(\theta), 0, \cos(\theta))$, and $R_\psi^{(0,0,1)}$ denotes the rotation matrix corresponding to the random rotation of angle $\psi$ around the CVAE-reference direction $\omega_\text{in}^{\text{CVAE}} = (0,0,1)$.

Given $x$, a local frame of reference can be obtained in $x$ by considering the coordinate system
\begin{equation*}
\hat{e}_\text{n} = \hat{x}\text{,}\quad \hat{e}_\text{b} = \omega_\text{in} \times \hat{x}\text{,}\quad
\hat{e}_\text{t} = \hat{e}_\text{b} \times  \hat{e}_\text{n}\text{,}
\end{equation*}
with $\hat{x} = x / r_\text{sphere}$, and $\hat{e}_\text{n}$, $\hat{e}_\text{b}$ and $\hat{e}_\text{t}$ referring to normal, binormal and tangent vectors of the frame of reference around $x$. The operator $\times$ denotes the cross product in $\mathbb{R}^3$. Within this system, the out-going direction $\omega$ can be parameterized through a tuple $(\alpha, \beta)$, wherein $\alpha = \cos(\omega, \hat{e}_\text{b})$ and $\beta = \cos(\omega, \hat{e}_\text{t})$. Again exploiting rotation invariance, $\omega$ can be sampled in the CVAE-reference system and can be transformed to real-space by applying a transformation similar to that in Equation~\ref{eq:transform}.

The treatment of summary statistics $\Sigma$ has to be considered separately and depending on which quantities are required. For the case of $\Sigma = (X, W)$, with $X$ and $W$ describing position and direction of a representative scattering event, a symmetry-guided restriction of sample space is not admissible since the statistics may depend on $x$ and $\omega$ in a complex fashion. Therefore, $X$ and $W$ are sampled as real-valued 3D vectors in the CVAE system of reference with $W$ being normalized subsequently. Both quantities are then transformed to real-space coordinates by applying a suitable rotation matrix and scaling.

To account for varying radii $r_\text{sphere}$ in the model setting, the extinction coefficient of the medium is rescaled according to $\sigma_\text{t} \rightarrow \sigma_\text{t} r_\text{sphere}$ before applying the models. This is justified because of homogeneity of the medium and independence of scattering events therein. All other variables are unaffected by the transformation between model- and real-space coordinates. 





\subsection{Model details and shader implementation}


Building upon the above considerations, the statistical model described above can be decomposed into a small number of shader functions, which accept input parameters according to the probabilities defined in Equation~\eqref{eq:probabilities} and sample output parameters as discussed in Section~\ref{sec:pathparam}. Three of the shader functions involve CVAE-based sub-steps and can be summarized follows:
\begin{itemize}
	\item Sampling of the number of scattering events:
	\begin{equation*}
	\textsc{LengthGen}(\sigma_t, g, z^{(L_\text{L})}, \epsilon^{(1)}) \rightarrow N\text{.}
	\end{equation*}
	\item Sampling of outgoing position and direction:
	\begin{equation*}
	\textsc{PathGen}(\sigma_t, g, N, z^{(L_\text{P})}, \epsilon^{(3)}) \rightarrow (\theta,\beta,\alpha)\text{.}
	\end{equation*}
	\item Sampling the representative scattering event:
	\begin{equation*}
	\textsc{EventGen}(\sigma_\text{t}, g, \varphi, \theta, \alpha, \beta, N, z^{(L_\text{E})}, \epsilon^{(6)}) \rightarrow (X, W)\text{.}
	\end{equation*}
\end{itemize}
Therein, $z^{(L)} \sim \mathcal{N}(0, \mathbb{I}_L)$ denotes the Gaussian random variables used for sampling the CVAE priors, respectively, and $\epsilon^{(P_\text
	out)} \sim \mathcal{N}(0, \mathbb{I}_{P_\text{out}})$ denote Gaussian random variables used for sampling the decoder posterior, as discussed in Section~\ref{sec:cvaedescribtion}. The settings of hyper parameters $D$, $W$ and $L$ used for setting up the CVAEs are summarized in Table~\ref{tab:params}.

\begin{table}[!h]
	\caption{Examined and final parameter settings for the CVAE models \textsc{LengthGen}, \textsc{PathGen} and \textsc{EventGen}. Final parameter settings are highlighted. }
	\label{tab:params}
	\centering%
	\def\arraystretch{1.2}
	\begin{tabular}{|l|c|c|c|c|c|}
		\hline
		\ & $P_\text{in}$ & $P_\text{out}$ & $D$ & $W$ & $L$ \\ \hline
		\textsc{LengthGen} & 2 & 1 & 1, \underline{2}, 3 & 4, \underline{8}, 12 & \underline{2}, 6\\
		\textsc{PathGen} & 3 & 3 & 1, \underline{2}, 3 & 8, 12, \underline{16} & \underline{5}, 9\\
		\textsc{EventGen} & 7 & 6 &  1, \underline{2}, 3 & 12, \underline{16}, 20 & \underline{5}, 9\\
		\hline
	\end{tabular}%
\end{table}

\subsection{In-scattered direct illumination\label{sec:directillum}}



In the following, the inclusion of next-event estimation for in-scattering from direct illumination \cite{coveyou1967adjoint, novak2018monte} is described as a concrete application of the third model, \textsc{EventGen}. In standard path tracing, this information is either neglected during the internal scattering process and considered only at the position where the path exits the volume, or the in-scattering is considered at every internal sampling point, e.g., by using pre-computed photon maps~\cite{novak2012virtual} or direct illumination using next-event estimation approaches~\cite{coveyou1967adjoint,koerner2016subdivision, hanika2015manifold}. Since at render time, the samples at which internal scattering occurs are not available any more, we set in this case $\Sigma = (X, W)$, wherein $X$ indicates the position of the scattering event in the interior of the sphere and $W$ the direction of the ray arriving to $X$. At training time, $(X, W)$ is sampled from the set of all scattering events on the path through the sphere with weights according to the contribution of the in-scattering events to the luminosity of the ray. 

For example, the direct light contribution from a directional light source at position $x_\text{L}$ with power $\Phi$ along a path inside a translucent medium can be considered at every internal sampling point of the ray path. Characterizing the ray through the occurring scattering events, $(x_1, \omega_1), \ldots, (x_N, \omega_N)$, the total in-scattering $L$ due to direct illumination along the path is obtained as
\begin{equation}
\label{eq:directLighting}
L = \Phi \sum_{i=1}^n \varphi^i \, \rho_g(\omega_i, \omega_{x_\text{L}\rightarrow x_i}) \, \gamma(x_\text{L}\rightarrow x_i)\text{.}
\end{equation}
Here, $\omega_{x_\text{L}\rightarrow x_i}$ indicates the direction of the incoming direct illumination at position $x_i$ and $\gamma(x_\text{L}\rightarrow x_i)$ denotes the extinction of the light when traveling from the light source through the volume to the scattering position. 

Since in path tracing typically the expectation value $\langle L \rangle$ over many paths is considered, 
our approach is to approximate $L$ in Monte Carlo fashion by choosing one single representative sample $(X, W)$ per path and considering the in-scattering from the direct light source only for this scattering event. A sample $(X, W)$, thereby, is chosen to be the $k$-th scattering event in a path $\{(x_i,\omega_i)\}_{i=1}^N$ with probability $\varphi^k / \Lambda^{(N)}$, where $\Lambda^{(N)} = \sum_{i=0}^N {\varphi^i}$. This approximation does not alter $\langle L \rangle$ if a sufficient number of rays is considered in the average.

The crucial step in applying this approximation lies in efficiently computing the incoming direction of the direct light, $\omega_{x_\text{L}\rightarrow X}$, as well as the attenuation coefficient $\gamma(x_\text{L}\rightarrow X)$ for all positions $X$. In absence of refractive boundaries, the expressions simplify to
\begin{equation*}
\omega_{x_\text{L}\rightarrow X} = \frac{x_\text{L} - X}{\lVert x_L - X \rVert}
\quad \text{and} \quad
\gamma(x_\text{L}\rightarrow X) = \exp(-d_t(x_\text{L}\rightarrow X) \sigma_t)\text{,}
\end{equation*}
with $\lVert\cdot\rVert$ indicating the standard Euclidean vector norm and $d_t(x_\text{L}\rightarrow X)$ referring to the distance along the direct path between $x_\text{L}$ and $X$, during which the light travels through the medium. In the more general case of refractive medium boundaries being present, however, alternative computation methods have to be employed, which are able to account for refraction appropriately. A number of methods have been proposed to tackle the problem, including \cite{walter2009single,holzschuch2015accurate, koerner2016subdivision}. However, the choice of a specific method does not affect applicability and performance gain of the path-tracing acceleration method proposed here.

\section{Results\label{sec:results}}

\begin{figure*}
	\centering
	\includegraphics[width=.95\textwidth]{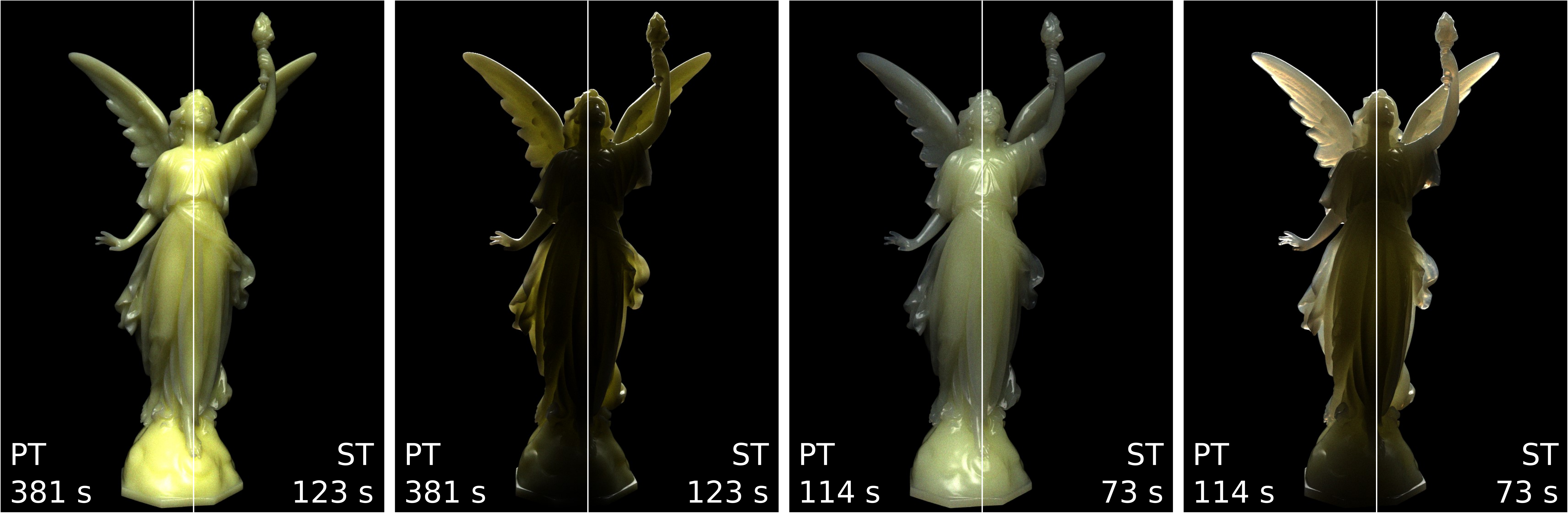}
	\caption{Comparison between path tracing (PT) and our sphere tracing approach (ST). From left to right: High-density material under front lighting and back-lighting conditions ($\sigma_\text{t}=(0.512\,\text{mm}^{-1}, 0.614\,\text{mm}^{-1}, 0.768\,\text{mm}^{-1})$ and $\varphi=(0.99999, 0.99995, 0.975)$ for channels $(\text{red}, \text{green}, \text{blue})$), and low-density material ($\sigma_\text{t}=(0.128\,\text{mm}^{-1}, 0.1536\,\text{mm}^{-1}, 0.192\,\text{mm}^{-1})$) under front-lighting and back-lighting conditions. All images were rendered using 5000 paths per pixel. Rendering times are shown in the images.}
	\label{fig:renderingcomparison}
	\end{figure*}
	
	To demonstrate the validity of our approach, we conducted numerical experiments regarding fitting accuracy of the model predictions with respect to the real distributions, visual accuracy of the rendering results including and excluding the in-scattered direct illumination, as well as performance analysis of the method in several scenarios. Note here that for all scenarios the same model configuration was used, i.e., no models were retrained to specialize for particular tasks. Training of the models was conducted on $1.6\times10^6$ ray samples with material parameters sampled from the ranges $\sigma_\text{t}\in[0, 200]$ and $g\in [-1, 1]$ to guarantee an appropriate coverage of material parameters relevant for our study. For all our experiments we assume model height to be normalized to $1\text{m}$.
	
	\subsection{Visual accuracy}
	
	Figure~\ref{fig:renderingcomparison} shows a direct comparison between renderings computed with our proposed sphere-tracing approach and with a brute force path tracer. In-scattered direct illumination is not yet included at this stage, so only \textsc{LengthGen} and \textsc{PathGen} are employed for rendering. To illustrate the capabilities and limitations of our method, we compare the same geometric model in two density configurations (high density and low density) and under different lighting conditions (light coming from a single light source in front of and behind the object). Since both approaches perform ray tracing and, thus, compute the same quantities, a fair comparison is achieved by fixing the number of rays traced per pixel (5000 in the current setting) and comparing the results and rendering times. 
	
	Figure~\ref{fig:renderingcomparison} reveals hardly any perceptual differences between our approach and brute force path tracing. Quantitatively, our renderings achieve a root mean square error to the path-traced ground truth between $0.01$ and $0.02$ and, thus, come very close to the ground truth. Especially the low-density rendering under backside lighting demonstrates that our approach is well able to generate translucency effects in thin material parts, which closely resemble the effects observed in the brute-force rendering. 
	
	\subsection{Performance analysis}
	
	\begin{figure*}
		\centering
		\includegraphics[width=.9\textwidth]{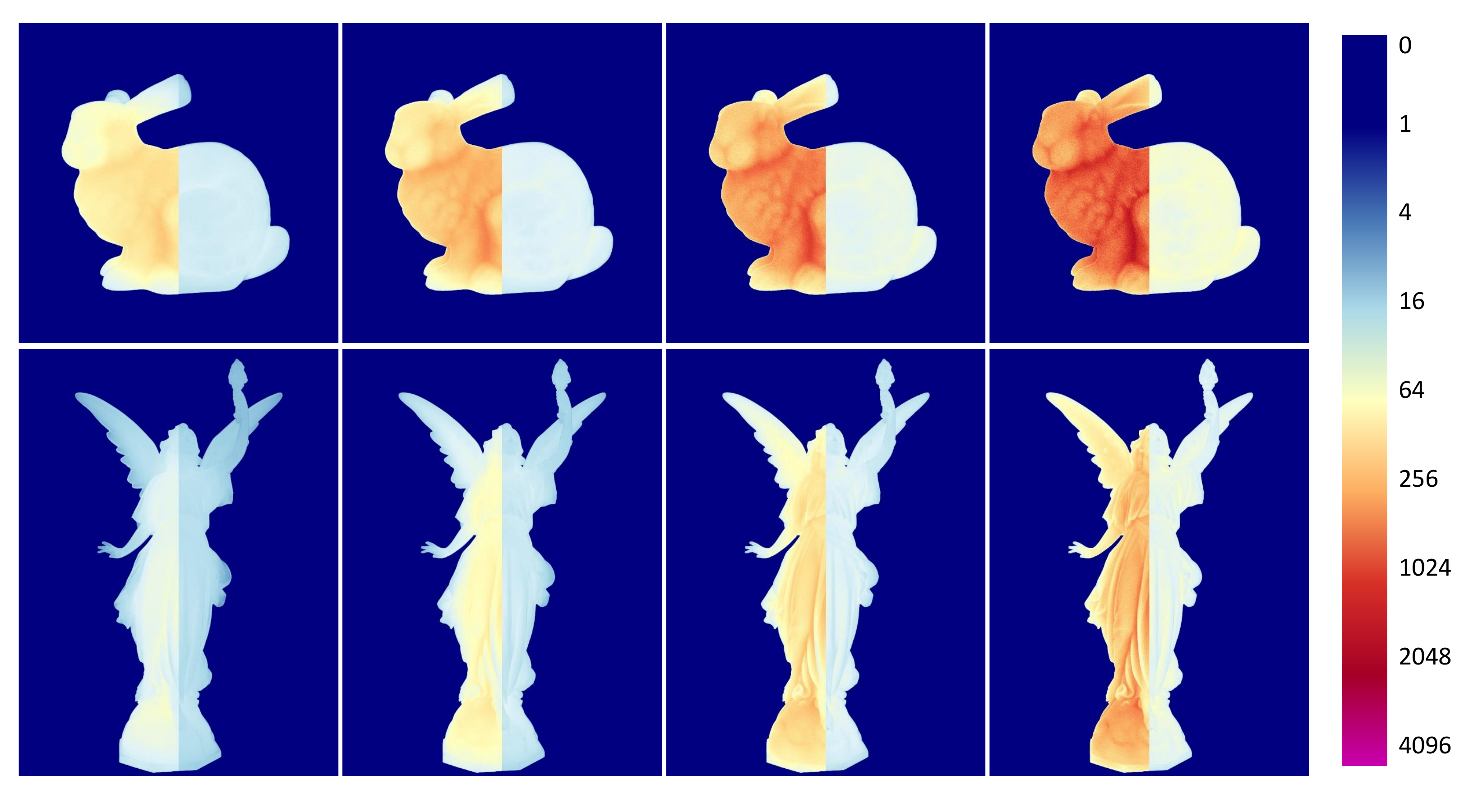}
		\caption{Number of scattering events per path for different geometries and densities. Standard path tracing (left) vs our approach (right). Densities increase from left to right.}
		\label{fig:numscatters}
	\end{figure*}
	
	\begin{figure}
		\centering
		\includegraphics[width=\columnwidth]{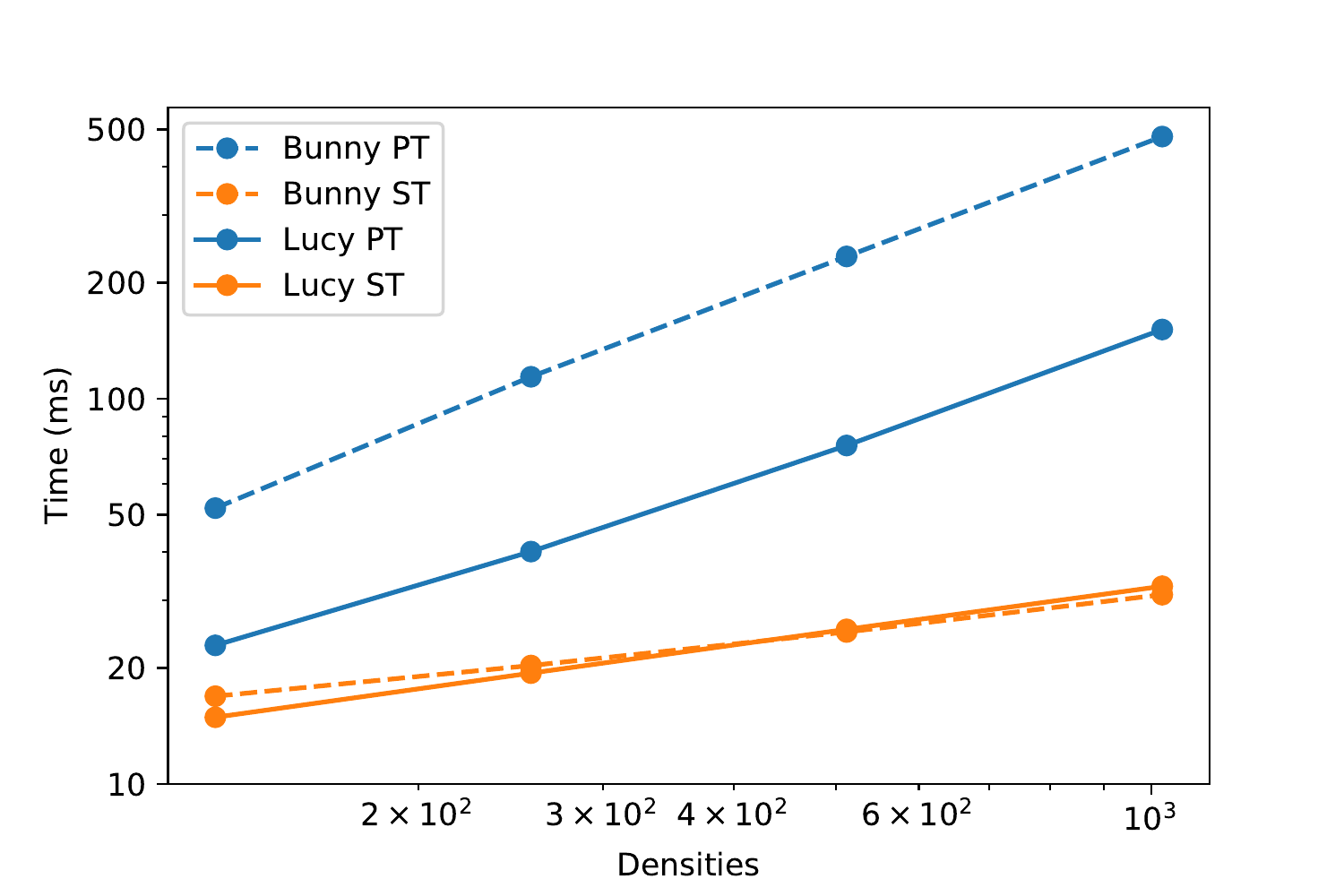}
		\caption{Time complexity of our approach (sphere tracing, ST) vs. brute-force path tracing (PT) for different geometries and optical densities.}
		\label{fig:timecomplexities}
	\end{figure}
	
	In the low-density setting in Figure~\ref{fig:renderingcomparison}, our approach achieves a speed-up of roughly $1.5$x ($73\,\text{s}$ for sphere tracing vs. $114\,\text{s}$ for path tracing) compared to brute force path tracing. The speed-up increases with material density, so that in the high-density case our approach is already $3.1$x faster than path tracing ($123\,\text{s}$ for sphere tracing vs. $381\,\text{s}$ for path tracing). This trend continues when the optical density of the material is further increased. As seen in the teaser (Figure~\ref{fig:teaser}), speed-ups of roughly $7.7$x or more can be achieved for sufficiently dense materials. 
	
	Given the increase in relative performance gain of our method with respect to standard path tracing, we further shed light on the different factors causing this behaviour. Therefore,  we rendered different geometric objects with varying parameter settings and examined the number of scattering events per path as well as the rendering performance depending on object geometry and optical material density. Figure~\ref{fig:numscatters} visually encodes the number of scattering events occurring per path for different geometries. The corresponding rendering times of our approach and brute force path tracing are shown in Figure~\ref{fig:timecomplexities}. 
	
	The Stanford bunny in the top row of Figure~\ref{fig:numscatters} is characterized by a compact shape, with many convex parts. As a result, light can penetrate deeply into the body, yielding long light paths with a large number of scattering events. As to be expected, the number of scattering events per path increases strongly with increasing optical density in brute force path tracing (left). Our method, in contrast, can take advantage of the voluminous convex shape to construct large spheres, which efficiently propagate the rays through the medium. Even though also in our approach the number of scattering events increases with increasing optical density, the rate is much lower. While in the case of the highest density the brute force path tracer is required to simulate up to $3000$ scattering events per path, our method propagates the rays in less than 50 steps. The need for sequential path evaluations, which cannot be parallelized on the GPU, is thus reduced by a factor of up to $60$x, and drastically reduces the amount of time that is spent waiting for long sequences of scattering events to converge. The rendering times in Figure~\ref{fig:timecomplexities} confirm these findings. Notably, compared to our approach the time increases at a higher rate when using brute force path tracing. In the case of highest optical density, a rendering time for a per-pixel path-trace of about $30\,\text{ms}$ stands against a rendering time of more than $500\,\text{ms}$, yielding a speed-up by a factor of more than $16$x. The difference between the reduction in scattering events, $60$x, and the final speed-up can be attributed to the increasing cost of evaluating the neural network models, compared to the few operations required in a standard path tracing.
	
	When rendering the statue model in the bottom row of Figure~\ref{fig:numscatters}, however, some limitations of our approach become paramount. Since the statue possesses many thin features, which do not form a voluminous material body, the sphere tracing approach is unable to build large spheres and propagate rays efficiently. Especially in the case of low-densities, our approach requires only slightly less model evaluations than the brute force method. In combination with comparatively costly model evaluations in our approach, the performance gain becomes very low, yielding a speed-up of less than $2$x (see Figure~\ref{fig:timecomplexities}). Nevertheless, for larger material densities the number of scattering events per path increases much stronger in the brute force method than in our approach, so that again a significant performance gain can be achieved, i.e. roughly $5$x at the highest density.
	
	\begin{figure}
		\centering
		\subfloat[Geometric model, rendered with different absorption albedos, using brute force path tracing (left half images) and our approach (right half images). The initial absorption albedo (left-most image) is $(2\times10^{-3}, 2\times10^{-4}, 2\times10^{-3})$ and is doubled between subsequent images.\label{fig:absorptionimages}]{    \includegraphics[width=\columnwidth]{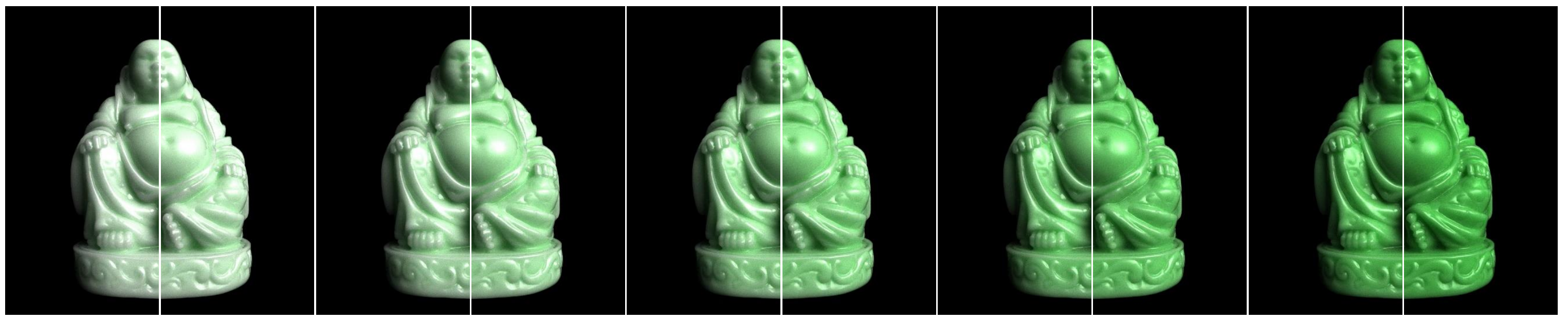}}\\
		\subfloat[Rendering times for the model from Figure~\ref{fig:absorptionimages} as a function of the absorption parameter.]{    \includegraphics[width=\columnwidth]{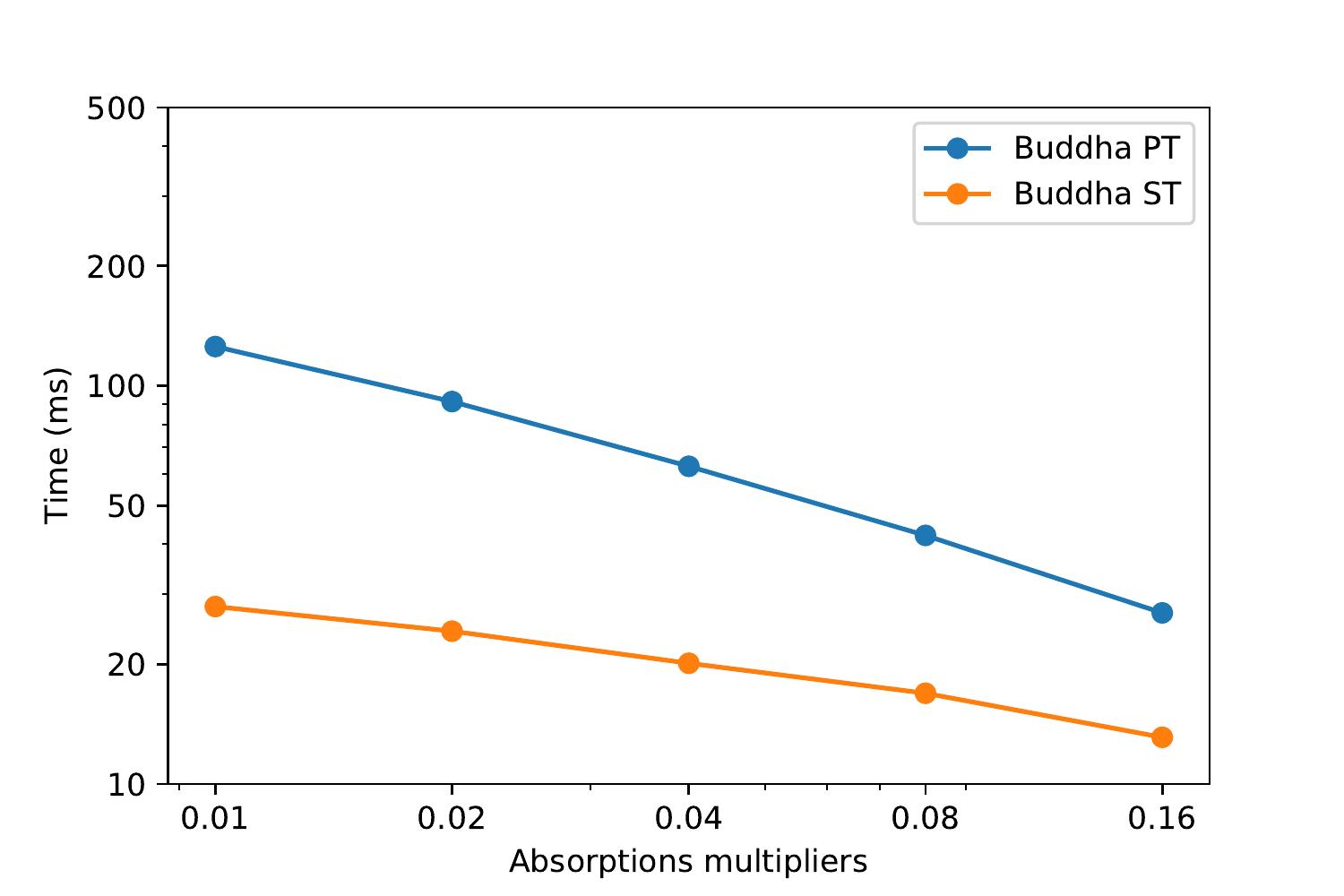}}
		\caption{Performance impact of the absorption parameter.}
		\label{fig:absorptions}
	\end{figure}
	
	From these considerations, we conclude that our method can be employed at high efficiency if standard path tracing leads to exceedingly long scattering paths. Thus, another factor limiting  efficiency of our approach is absorption, which has not been considered strongly in previous experiments. Figure~\ref{fig:absorptions} explores this effect in detail. For this experiment, we rendered a geometric object with constant material parameters and under fixed lighting conditions, while varying the absorption rate of the medium ($1 - \varphi$). From the figure it is obvious that higher absorption leads to shorter ray paths and thus shortens the rendering time. At higher absorption, the advantage of our approach decreases.

	\subsection{Quality of the learned distributions}
	
	To evaluate the accuracy of plain model predictions, we conduct the following experiment. For a predefined set of material parameters, $\sigma_\text{t} \in \{1, 4, 20, 100\}$, $g\in\{-0.7, 0.0, 0.4, 0.9\}$, $\varphi=1$, we use standard path tracing to generate a distribution of ground-truth rays, starting in the center of a unit sphere in direction $\omega_\text{in} = (0, 0, 1)$ and evaluate the ray statistics $(N, \theta, \alpha, \beta, X, W)$ for each of the rays.
	The resulting distribution is then compared to the distribution, which is obtained from sampling the respective CVAE models. For all experiments, we draw $10000$ samples per parameter setting $(\sigma_\text{t}, g)$, for both ground truth and CVAE statitics..
	
	\begin{figure}
		\centering
		\includegraphics[width=.95\columnwidth]{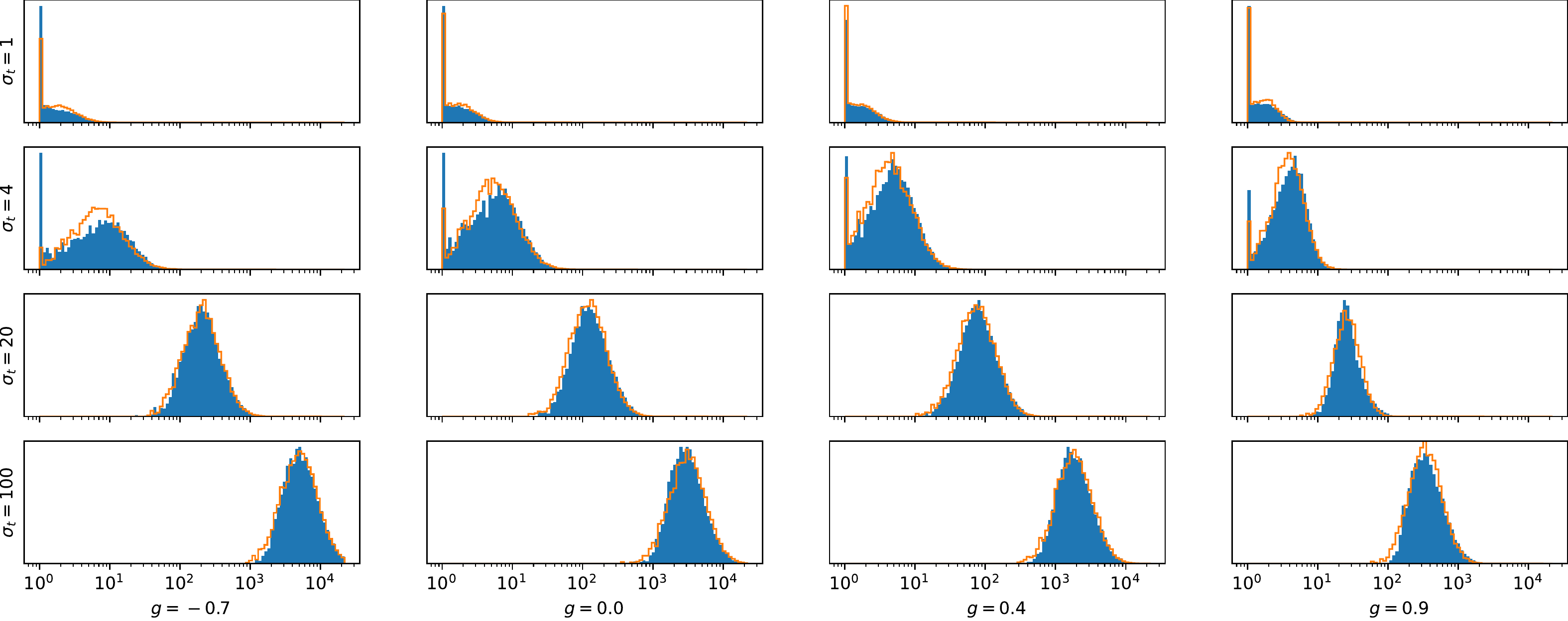}
		\caption{Distribution of number of samples $N$ from \textsc{LengthGen} (orange) vs. ground truth (blue) for different settings of $\sigma_t$ and $g$.}
		\label{fig:ndistributions}
	\end{figure}
	
	The distribution of $N$ samples drawn from the \textsc{LengthGen} model is compared to the respective ground truth in Figure~\ref{fig:ndistributions}. Ground truth statistics are summarized in a log-scale histogram, shown in blue, and overlaid by an orange curve reflecting the corresponding statistics obtained from sampling \textsc{LengthGen}. It can be seen that for small values of $\sigma_\text{t}$ (top row), a large fraction of rays leaves the sphere already after the first scattering event. This results in a bipartite structure of the histogram, consisting of a pronounced peak of the histogram counts around $N=1$ and a smooth tail indicating rays that are scattered more than once. \textsc{LengthGen} learns to compensate for this and is able to reproduce the peak accurately, especially in the case of strongly anisotropic forward scattering, $g=0.9$. Only in the case of backward scattering, $g=-0.7$, the ground truth is underestimated significantly. Nevertheless, the distribution of scatter counts in the case of $N>1$ is reproduced to reasonable accuracy in all cases. For larger values, the significance of the peak at $N = 1$ decreases and the ground-truth statistics appear to converge to a unimodal shape, similar to a log-Gaussian. For larger values of $\sigma_\text{t}$, notably $\sigma_\text{t} = 20, 100$, the matching of the distributions is good, and also the dependence on $g$ is captured appropriately. The highest inaccuracy is observed at intermediate scattering coefficients, $\sigma_\text{t} = 4$, where the multi-scattering contributions are dominant, but single scattering still has a notable contribution. There, it seems difficult for the model to predict the balance between both effects, especially as a function of $g$. These observations can be used, however, to bias the statistics of $\sigma_\text{t}$ and $g$. By increasing the number of samples in these difficult regions a more accurate fit can be achieved. 
	
\begin{figure}
		\centering
		\subfloat[Distribution of $\cos(\theta)$.\label{fig:thetadistribution}]{
			\includegraphics[width=.95\columnwidth]{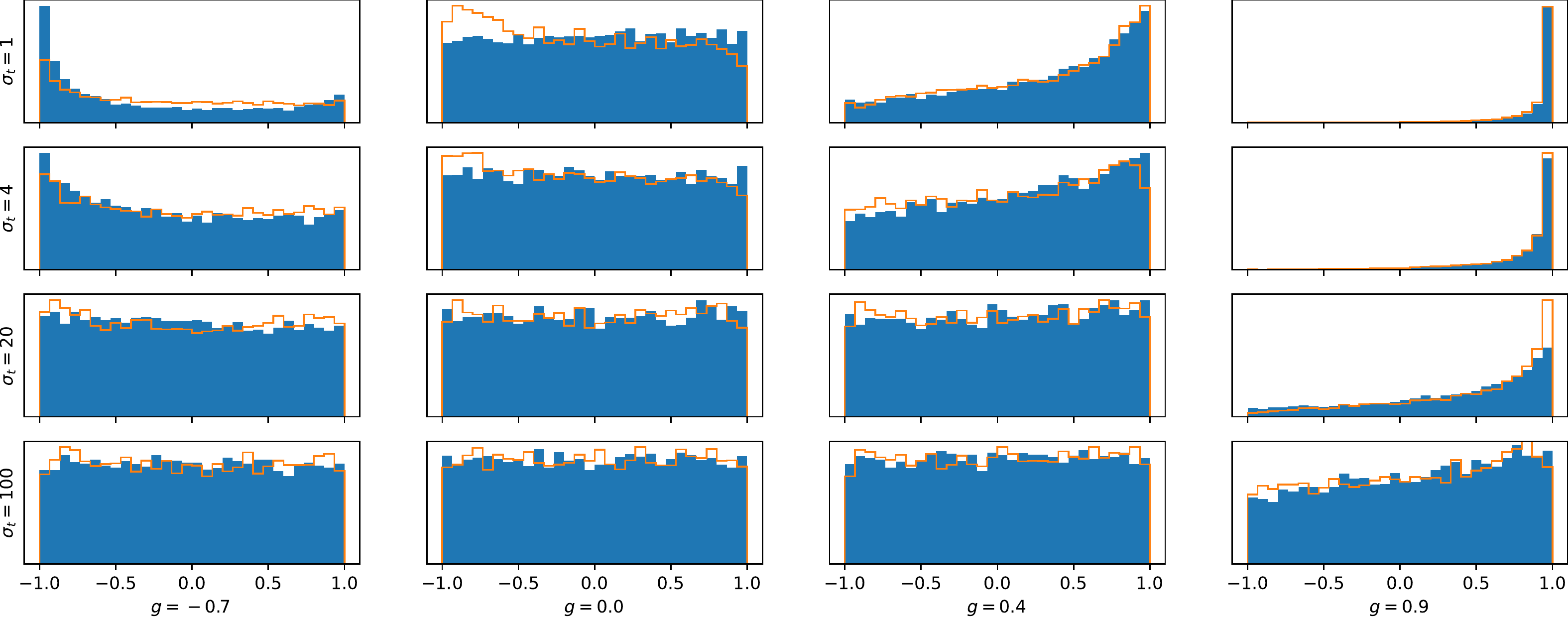}
		}\\
		\subfloat[Distribution of $\alpha$ (vertical axis) as a function of $\cos(\theta)$ (horizontal axis).\label{fig:adistribution}]{
			\includegraphics[width=.45\columnwidth]{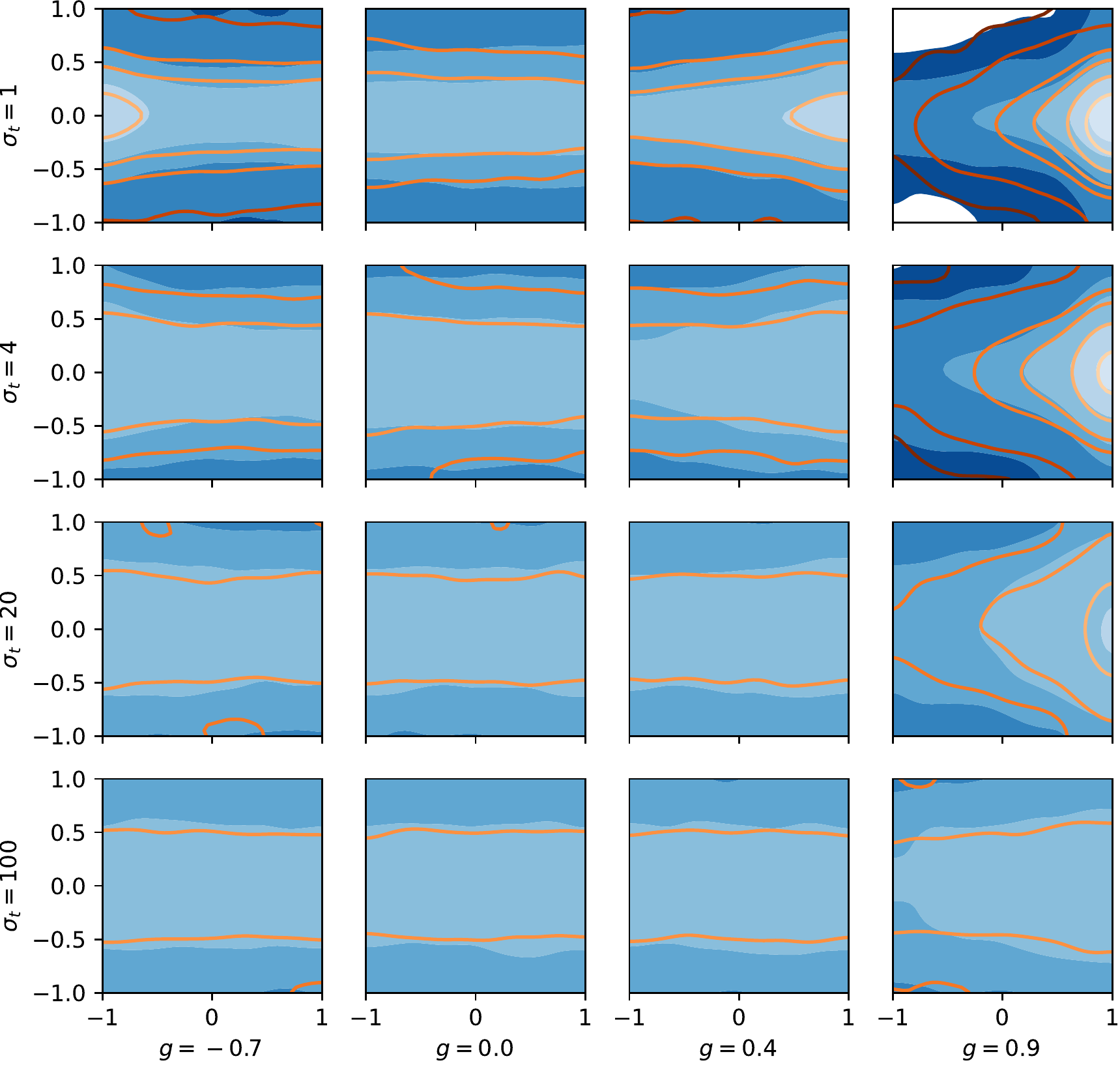}
		}
		\hfill
		\subfloat[Distribution of $\beta$ (vertical axis) as a function of $\cos(\theta)$ (horizontal axis).\label{fig:bdistribution}]{
			\includegraphics[width=.45\columnwidth]{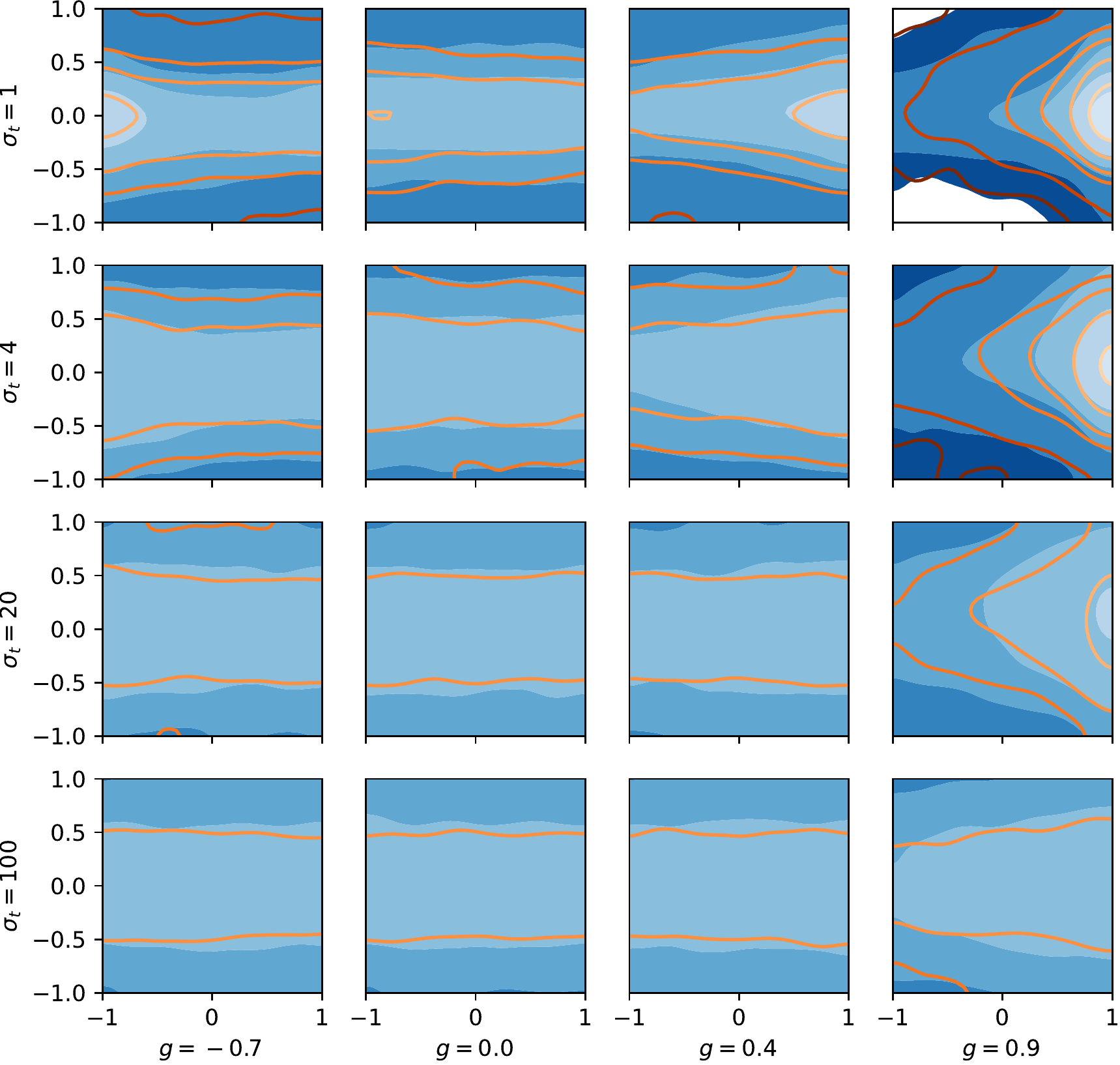}
			
		}
		\caption{Distribution of \textsc{PathGen} samples for  $\cos(\theta)$, $\alpha$ and $\beta$ (orange) vs. ground truth (blue) for different settings of $\sigma_t$ and $g$.}
	\end{figure}

	For the \textsc{PathGen} model, three parameters need to be validated against ground truth. The sampling statistics of $\cos(\theta)$ are shown in Figure \ref{fig:thetadistribution}. Again, the blue histogram refers to ground truth statistics, whereas the orange line reflects the \textsc{PathGen} predictions. Notably, strongly anisotropic scattering is observed only in cases where $\sigma_\text{t}$ is small or $g$ is close to extreme values $-1$ or $1$. There is an overall good fit between learned and ground-truth distributions, only for strongly anisotropic cases, $g = 0.9$ and $g=-0.7$, the model has minor difficulties with correctly handling the importance of single-scattering rays. 
	
	To validate the distribution of $\alpha$ and $\beta$, we generate a contour map, which shows the distribution function of $\alpha$ and $\beta$ as a function of $\cos(\theta)$. Two-dimensional histograms of the distribution of $(\cos(\theta), \alpha)$ and $(\cos(\theta), \beta)$ are obtained and normalized by the number of observations. The resulting bin probability values are finally smoothed with a Gaussian filter to reduce visual clutter. The ground-truth distribution is shown as blue-filled contours, the learned distribution is overlaied as orange isocontours at probability iso-values matching those of the ground truth. The comparison is shown in Figures~\ref{fig:adistribution} and~\ref{fig:bdistribution}. Similar to the distribution of $\cos(\theta)$, the greatest anisotropy is observed for highly anisotropic cases, i.e., $g=0.9$ or $g=-0.7$, and small $\sigma_\text{t}$. Nevertheless, the learned distributions fit the ground truth to reasonable accuracy.
	
	We conclude that the major challenges for all models lie in learning a reasonable PDF at small values of $\sigma_\text{t}$ and in correctly resolving the dependence of the PDF on $g$. From a practical perspective, learning results can be improved, however, by specializing the model at training time towards those parameter regions, which are particularly relevant during upcoming rendering processes. By over-weighting ray samples with particular $(\sigma_\text{t}, g)$-settings at training time, these regions of parameter spaces are assigned a larger weight in the loss function and thus affect the outcome of the training process more strongly. If, for example, only media with largely forward-directed scattering are considered during rendering, there is no need to include negative values of $g$ at training time. For our purposes, however, we did not make use of such optimizations.
	
	\subsection{In-scattered direct illumination}
	
	\begin{figure*}
		\centering
		\includegraphics[width=\textwidth]{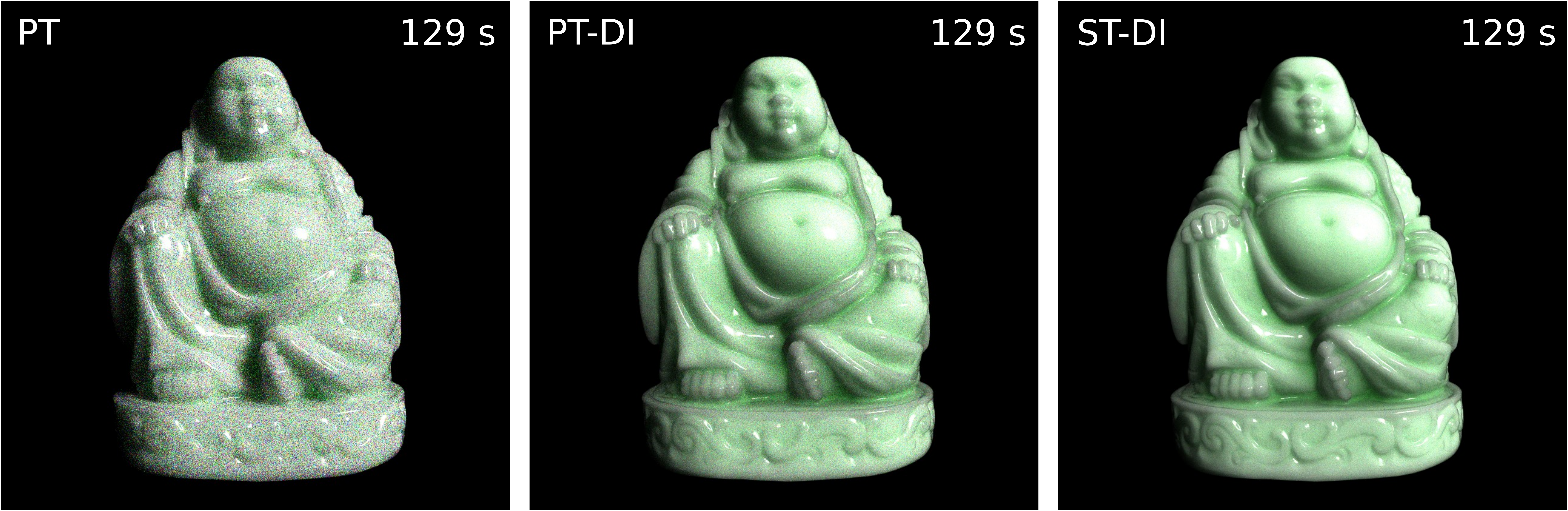}
		\caption{Convergence improvement using next-event estimation. All images are rendered in $129\,\text{s}$. Left: Brute-force path tracing (1000 rays per pixel), excluding next event estimation. Middle: Brute-force path tracing (400 rays per pixel), including next-event estimation using in-scattered direct illumination at every internal sample. Right: Our approach (1465 rays per pixel), including next-event estimation using one representative sample ($X$,$W$) per sphere tracing step.}
		\label{fig:directin}
	\end{figure*}
	
	To also illustrate the use of the third CVAE model, \textsc{EventGen}, we compare the result of rendering a geometric object with the full sequence of three models, \textsc{LengthGen}, \textsc{PathGen} and \textsc{EventGen}. For comparison purposes, we render the same image also with a standard path tracer, as well as with a brute-force path tracer with next-event estimation using direct illumination. To obtain a fair comparison, the image was first rendered with the standard path tracer, using 1000 ray samples per pixels and rendering time was measured to be $129\,\text{s}$. The remaining renderers finally were allowed to render the same geometry for the same amount of time.
	In the current implementation, both the extended brute-force path tracer as well as the sphere tracing model draw a shadow ray to a single light source in the scene setting to account for direct illumination. As such, they do not handle refractive object boundaries accurately and yield biased results in comparison with the standard path tracer. Nevertheless, Figure~\ref{fig:directin} illustrates nicely that our methods yields a lower variance than both other algorithms. This is a promising result and suggests that CVAEs can be employed successfully for learning summary statistics of the bypassed path sections. Also, the performance gain achieved is orthogonal to the problem of accurately treating refractive boundaries, since potential fixes, enabling accurate inclusion of boundary effects, would affect our model and the reference path tracer in identical ways.

\section{Conclusion}

In this work we have introduced and analyzed a network pipeline comprised of CVAEs that learns to bypass multiple-scattering paths in anisotropic volumetric material. We have demonstrated that this pipeline can be embedded into a sphere-tracing approach to speed up the performance of Monte Carlo path tracing. For different parameter settings and geometries, we have demonstrated numerical accuracy of the inferred results. 
We are particular intrigued about the visual quality of the rendering results compared to ground truth Monte Carlo path tracing, as well as the performance gain that can be achieved for objects with large volumes of material. 

On the other hand, we have seen that for thin objects where sphere tracing cannot exploit its full potential, the performance gains are only marginal. To minimize the overhead in computation time in low-density or thin object parts, we will consider a case-sensitive implementation of the sphere-tracing algorithm in the future. For shell tracing ~\cite{moon2007rendering}, a fall-back strategy to standard path tracing was already used to avoid sphere tracing in the limit of very small radii. Efficiently parallelizing CVAE evaluation and standard path tracing on GPU architectures concurrently, however, appears challenging. Furthermore, the proposed combination of deep learning and sphere tracing opens the possibility to efficiently render volumetric objects like clouds. In this context, we envision, in particular, the integration of direct in-scattering to result in significantly improved performance. 




\bibliographystyle{abbrv}
\bibliography{refs}
\end{document}